\documentclass[fleqn,usenatbib]{mnras}

\usepackage{newtxtext,newtxmath}

\usepackage[T1]{fontenc}
\usepackage{ae,aecompl}

\usepackage{graphicx}	
\usepackage{amsmath}	
\usepackage{amssymb}	
\usepackage{aas_macros}
\usepackage{hyperref} 
\hypersetup{colorlinks,citecolor=blue,linkcolor=blue,urlcolor=blue}
\usepackage{gensymb}
\usepackage{CJKutf8}
\usepackage{subfig}
\usepackage{arydshln}
\usepackage{threeparttable}

\newcommand{\mdmunits}{{\rm pc \, cm^{-3}}} 
\newcommand{\dmunits}{$\mdmunits$}

\title[Time Resolved Studies of 5 ASKAP FRBs]{High time resolution and polarisation properties of ASKAP-localised fast radio bursts}

\author[Cherie K. Day et al.]{Cherie K. Day$^{1,2}$\thanks{contact: cday@swin.edu.au}, 
A.~T.~Deller$^1$, 
R.~M.~Shannon$^1$, 
Hao~Qiu(\begin{CJK*}{UTF8}{gbsn}邱昊\end{CJK*})$^{3,2}$, 
\newauthor
Keith~W.~Bannister$^2$, 
S.~Bhandari$^2$, 
Ron~Ekers$^{2,4}$, 
Chris~Flynn$^1$, 
C.~W.~James$^4$,
\newauthor
J.-P.~Macquart$^{4\thanks{Deceased}}$,
Elizabeth~K.~Mahony$^2$,
Chris J. Phillips$^2$
and J.~Xavier~Prochaska$^{5,6}$
\\
$^{1}$Centre for Astrophysics and Supercomputing, Swinburne University of Technology, Hawthorn VIC 3122, Australia\\
$^{2}$Commonwealth Science and Industrial Research Organisation,  Australia Telescope National Facility, P.O. Box 76, Epping, NSW 1710 Australia\\
$^{3}$Sydney Institute for Astronomy, School of Physics, University of Sydney, NSW 2006, Australia\\
$^{4}$International Centre for Radio Astronomy Research, Curtin Institute of Radio Astronomy, Curtin University, Perth, WA 6845, Australia.\\
$^{5}$Department of Astronomy \& Astrophysics, UC Santa Cruz, USA\\
$^{6}$Kavli Institute for the Physics and Mathematics of the Universe (Kavli IPMU; WPI), The University of Tokyo, Japan
}

\date{Accepted XXX. Received YYY; in original form ZZZ}

\pubyear{2020}

\begin{document}
\label{firstpage}
\pagerange{\pageref{firstpage}--\pageref{lastpage}}
\maketitle

\begin{abstract}
Combining high time and frequency resolution full-polarisation spectra of Fast Radio Bursts (FRBs) with knowledge of their host galaxy properties provides an opportunity to study both the emission mechanism generating them and the impact of their propagation through their local environment, host galaxy, and the intergalactic medium.
The Australian Square Kilometre Array Pathfinder (ASKAP) telescope has provided the first ensemble of bursts with this information. In this paper, we present the high time and spectral resolution, full polarisation observations of five localised FRBs to complement the results published for the previously studied ASKAP FRB~181112. We find that every FRB is highly polarised, with polarisation fractions ranging from 80 -- 100\%, and that they are generally dominated by linear polarisation. While some FRBs in our sample exhibit properties associated with an emerging archetype (i.e., repeating or apparently non-repeating), others exhibit characteristic features of both, implying the existence of a continuum of FRB properties. When examined at high time resolution, we find that all FRBs in our sample have evidence for multiple sub-components and for scattering at a level greater than expected from the Milky Way. We find no correlation between the diverse range of FRB properties (e.g., scattering time, intrinsic width, and rotation measure) and any global property of their host galaxy. The most heavily scattered bursts reside in the outskirts of their host galaxies, suggesting that the source-local environment rather than the host interstellar medium is likely the dominant origin of the scattering in our sample.

\end{abstract}




\section{Introduction}
Fast radio bursts (FRBs) are bright, of order microsecond to millisecond duration bursts of radio emission that have been observed from from 300 MHz \citep{Chawla2020} to 8 GHz \citep{Hessels2019ApJ...876L..23H}. With observed peak flux densities in the range $\sim50$~mJy to $800$~Jy \citep{Petroff2019,Macquart_2019} and cosmological distances, their inferred luminosities are more than $12$ orders of magnitude brighter than the brightest regular pulsar pulses \citep{Macquart_2019}, pointing to an extreme and, as yet, unknown progenitor and emission mechanism.

The high time resolution, spectropolarimetric properties of FRBs are crucial to constraining both their emission physics and the local environments.
For instance, the $\rm \sim 30 \mu s$ microstructure observed by \citet{Farah18}  in FRB~170827 implies emission regions $\sim 10$~km in size, while the tens of microsecond sub-pulse structure reported by \citet{Cho2020arXiv200212539C} constrains the physical source size of FRB~181112 to a few kilometres.
The temporal evolution of the burst polarisation on comparable timescales also yields information on the emission process. \citet{Cho2020arXiv200212539C} inferred potential emission region and magnetic field topology in FRB~181112 based on the the variations in the burst polarisation position angle (PA). They found the burst comprised four distinct sub-pulses, and found not only a differential RM between sub-pulses but also a possible differential dispersion measure (DM), with the final sub-pulse exhibiting a residual delay in its frequency-arrival times. Moreover, the variation in the circular polarisation across the burst profile provided evidence that its radiation propagated through a relativistic plasma in the source region. While relatively few FRBs have polarisation information, similar circular polarisation changes have been observed in other FRBs \citep[e.g.,][]{Petroff15,Masui2015,Caleb2018MNRAS.478.2046C}, implying this might be a fairly common feature.

The propagation effects of Faraday rotation and plasma scattering likewise play a key role in diagnosing both the intervening and circumburst environments. Large scattering and RM magnitudes in FRBs have led to speculation that the circumburst environment of some FRB sources might be highly dense and magnetised \citep[e.g.,][]{Masui2015}.  However, while the $\rm |RM| \sim$~$10^5$~rad~m$^{-2}$ of FRB~121102 \citep{Michilli2018Natur.553..182M} indicates a dynamic, highly ordered, strong magnetic field near the source, it exhibits negligible scattering \citep[e.g.,][]{Hessels2019ApJ...876L..23H}. The RMs of all other bursts with detected linear polarisation are much less extreme: these range from no measurable RM at all \citep[e.g.,][]{Petroff15,Kumar2019ApJ...887L..30K} to a few to tens of $\rm rad \, m^{-2}$ \citep[e.g.,][]{Ravi1249,Petroff2017MNRAS.469.4465P} to a few hundreds of $\rm rad \, m^{-2}$ \citep[e.g.,][]{Masui2015,Caleb2018MNRAS.478.2046C}.
In addition, scattering and scintillation can yield clues to the characteristics of the material local to the source and intersected along the line of sight. Investigating the two distinct spectrotemporal modulation features observed in FRB~170827, \citet{Farah18} concluded they could be explained by the presence of two scattering screens, both resulting in scintillation of the burst. While the larger scale scintillation is consistent with that expected along the line of sight for a screen within the Galactic interstellar medium (ISM), the small-scale striations implied a second scattering screen within 60~Mpc of the source. Further constraints on the local environment, however, were hampered by the lack of a host galaxy identification.

The advent of localisation has transformed our ability to connect the spectropolarimetric properties of detected FRBs with their environments. The localisation of the repeating FRB~121102 to a high star formation rate region within a dwarf galaxy \citep{Tendulkar17,Chatterjee17} together with high time and frequency resolution, full polarisation data \citep[e.g., ][]{Michilli2018Natur.553..182M,Hessels2019ApJ...876L..23H} has facilitated an unprecedented wealth of information about the origins and surroundings of this FRB. \citet{Bannister565} reported the first localisation of a one-off burst, associating FRB~180924 with a massive, relatively quiescent galaxy, which cast doubt on FRB progenitor theories based on FRB~121102 that required prolific recent star formation. Subsequently, the localisation of FRB~181112 demonstrated the effectiveness of FRBs as cosmological tools. The intersection of the FRB~181112 sightline with the circumgalactic medium (CGM) of an intervening galaxy enabled stringent constraints on its halo gas density, magnetisation and turbulence to be derived from burst polarisation and high time resolution ($\rm 54 \mu s$) information \citep{Prochaska231}.

The higher quality data typically available for repeating FRBs have led to a number of insights regarding possible emission mechanisms \citep[see e.g., ][and references therein]{Platts_FRBtheoryCat}. While theories have often been tailored to FRB~121102, as it has been the most exhaustively studied, they have recently been challenged by subsequent localisations of as-yet non-repeating FRBs \citep{Bannister565,Ravi_2019,Prochaska231,Macquart2020_DMz} and a second localised repeating FRB \citep[FRB~180916.J0158+65][]{Marcote2020Natur.577..190M}. The full polarisation, higher time resolution data available for FRB~121102 \citep{Michilli2018Natur.553..182M} and FRB~180916.J0158+65 \citep{CHIME9repeaters2020ApJ...891L...6F} have also led to suggestions that polarisation properties might serve as a key discriminant of emission region characteristics between repeating and apparent non-repeating sources. Both are essentially 100\% linearly polarised and show a flat PA across their (wide) pulses \citep[][respectively]{Michilli2018Natur.553..182M,CHIME9repeaters2020ApJ...891L...6F}, contrasting the PA swings and circular polarisation seen in FRB~181112 \citep{Cho2020arXiv200212539C}. However, the comparative narrowness of most apparently non-repeating FRBs (and the lack of polarisation information in most cases) means that the constraints on the non-repeating population are much weaker.

In contrast to repeating FRBs, where the known position and DM facilitated the use of high time resolution recording systems \citep[e.g.,][]{Hessels2019ApJ...876L..23H}, apparently non-repeating FRB data quality is generally limited by the instrumental resolution of the FRB detector, which has historically suffered computational and data rate constraints. Until recently, only a few apparently non-repeating bursts have been detected in real time to trigger the storage of high-resolution data products that enable in-depth spectrotemporal property studies \citep[e.g.,][]{Farah18}

The capabilities of the Commensal Real-Time ASKAP Fast Transients (CRAFT) system on the Australian Square Kilometre Array Pathfinder (ASKAP) telescope, however, have recently allowed us to extend these studies to the population of apparently non-repeating FRBs \citep{Bannister565,Prochaska231,Cho2020arXiv200212539C}.
This offers the prospect of identifying key differences between these populations.

In this paper, we present the high time and frequency resolution, full polarisation results for five ASKAP-localised FRBs, forming a total sample of six exceptionally high signal-to-noise ratio, localised FRBs with spectropolarimetric information investigated at high time resolution \citep{Bannister565,Prochaska231,Macquart2020_DMz}. We examine their observed and derived properties in combination with their known hosts to form a collective picture of their properties and how these are correlated with their local and host galaxy environments, and we explore the potential distinctions between repeater-like and apparently non-repeater-like bursts. We describe the methods used to localise the bursts, calibrate their spectra, and extract the derived parameters in Section \ref{sec:method}. We provide an overview of the results in Section \ref{sec:results} and then proceed to discuss the characteristics of each FRB in Section \ref{sec:discussion}. Finally, Section \ref{sec:conclusions} explores the broader implications of the observed spectral, temporal, and polarimetric diversity within the FRB population.
\section{Methods} \label{sec:method}
The data acquisition for the ASKAP-CRAFT real-time detection system and the method used to determine the position and astrometric positional uncertainty of the FRBs in our sample follows that discussed in the Supplementary Materials (SM) of \cite{Bannister565}, \cite{Prochaska231}, and \cite{Macquart2020_DMz}. Briefly, three sets of dual linear polarisation, complex-sampled voltage data, 3.1~seconds in duration with a 336-MHz bandwidth, were captured for each FRB in our sample: the FRB, a phase and flux calibrator (a bright, compact radio source), and a polarisation calibrator (the Vela pulsar, PSR~J0835$-$4510). From these voltage data, the visibility datasets listed in Table~\ref{tab:vis_datasets} were made using the Distributed FX (DiFX) software correlator \citep{Deller11}.

The following is a general description of each visibility dataset:
\begin{itemize}
    \item \textit{FRB calibrator dataset}: the phase/flux calibrator data used to phase and flux calibrate all FRB datasets and the polarisation calibrator data. The full 3.1~s of data were correlated with the temporal and spectral resolutions given in Table~\ref{tab:vis_datasets}. PKS~0407$-$658 was used to calibrate FRB~180924, FRB~190611, and FRB~190711, while FRB~190102 and FRB~190608 were calibrated with PKS~1934$-$638. As outlined in the SM of \cite{Bannister565} and \cite{Prochaska231}, a clean portion of the total observing band (that is, one free from radio frequency interference [RFI]) was used to determine antenna-based, frequency-dependent delay solutions using the Astronomical Image Processing System \cite[\textsc{aips},][]{Greisen03} tasks \textsc{fring} and \textsc{calib}, which were subsequently applied to both the calibrator and target data. The \textsc{aips} task \textsc{cpass} was likewise used to correct for the instrumental bandpass.
    \item \textit{FRB position dataset}: the data used to determine the statistical position and uncertainty of the burst. These visibilities were made using the pulsar gating mode of DiFX, enabling the user to select the window of time (or ``gate'') in which the FRB signal is on and discard the remainder of the data. The optimal size of this gate depends on the duration of the pulse, and the temporal resolutions used for our sample are given in column 4 of Table~\ref{tab:vis_datasets}.
    \item \textit{FRB continuum field dataset}: the 3.1-s continuum background data used to align the ASKAP frame to the International Celestial Reference Frame \citep[ICRF3,][]{ICRF3_2018AGUFM.G42A..01G} and determine the astrometric uncertainties in the ASKAP data as outlined in \cite{Bannister565} and \cite{Prochaska231}. As with the calibrator data, the full 3.1~s of voltage data were integrated with the spectral and temporal resolutions listed in Table~\ref{tab:vis_datasets}.
    \item \textit{FRB HTR dataset}: the high time resolution (HTR) FRB data. The DiFX pulsar gating mode was used to correct for frequency-dependent dispersion and create multiple visibilities of a user-specified time resolution (see Table~\ref{tab:vis_datasets} column 4) that collectively span the duration of the FRB signal. We note that the DM taken from the detection was refined after inspection of initial HTR data, and the final correlation resulting in the reported \textit{FRB HTR} dataset used this optimised DM.
    \item \textit{Vela dataset}: the polarisation calibrator data (PSR~J0835$-$4510) used to correct the full Stokes spectra for each FRB dataset. As with the \textit{FRB position} data, the DiFX gating mode was used to isolate the Vela pulse, with the gate edges set to be roughly the burst width at 10\% of maximum intensity. See Section \ref{sec:pol_cal} for a description of the polarisation calibration.
    \item \textit{FRB (or Vela) (HTR) RFI subtraction dataset}: the data used to mitigate the RFI in either the FRB or Vela datasets. As with the target datasets listed above (\textit{FRB position}, \textit{FRB HTR}, and \textit{Vela}), these visibilities were created by correlating the target data in the DiFX pulsar gating mode. Here, however, they were correlated and integrated over a range of the data on either side of the target pulse, with a gap between the target gate edges and the two RFI gates in order to ensure none of the target signal would be removed. The total size of this RFI gate is given by the temporal resolution in Table~\ref{tab:vis_datasets} and is approximately symmetric about the target gate. As detailed in \cite{Bannister565} and \cite{Prochaska231}, a scaled version of the RFI subtraction visibility was subtracted from the target visibility using the custom \textsc{ParselTongue} \citep{Kettenis_ParselTongue2006ASPC..351..497K} script \textsc{uvsubScaled.py}, a task in the \textsc{psrvlbireduce} repository\footnote{\url{https://github.com/dingswin/psrvlbireduce}}. The RFI datasets were correlated with the same spectral resolution as their target counterparts. With the exception of the HTR datasets for FRB~190102, which reduced the correlation frequency resolution to 18.52 kHz in order to achieve  $54\mu \rm s$ temporal resolution, this was 9.26~kHz. All target datasets were RFI subtracted.
\end{itemize}
All datasets were further averaged in frequency after correlation by a factor of 27, resulting in resolutions of 250~kHz and 500~kHz for starting resolutions of 9.26~kHz and 18.52~kHz, respectively.
\begin{table*}
    \centering
    \begin{tabular}{lllll}
        \hline
        FRB & visibility dataset & correlation centre (R.A., Decl.) & temporal resolution & spectral resolution \\
         & & (J2000 hh:mm:ss.s, dd:mm:ss.s) & (sec) & (kHz) \\ \hline \hline
        FRB~180924 & \textit{FRB calibrator} & 04:08:20.38, $-$65:45:09.08 & 1.3824 & 9.26 \\
         & \textit{FRB position} & 21:44:25.2943, $-$40:53:59.9959 & 0.001 & 9.26 \\
         & \textit{FRB continuum field} & 21:45:17.83, $-$41:03:34.67 & 1.3824 & 9.26 \\
         & \textit{FRB HTR} & 21:44:25.2943, $-$40:53:59.9959 & 0.000108 & 9.26 \\
         & \textit{Vela} & 08:35:20.61149, $-$45:10:34.8751 & 0.009 & 9.26 \\
         & \textit{FRB RFI subtraction} & 21:44:25.2943, $-$40:53:59.9959 & 0.033 & 9.26 \\
         & \textit{Vela RFI subtraction} & 08:35:20.61149, $-$45:10:34.8751 & 0.030 & 9.26 \\ \hline
        FRB~190102 & \textit{FRB calibrator} & 19:39:25.0262814, $-$63:42:45.624366 & 1.3824 & 9.26 \\
         & \textit{FRB position} & 21:29:39.70836, $-$79:28:32.2845 & 0.001 & 9.26 \\
         & \textit{FRB continuum field} & 21:32:32.623, $-$79:17:18.38 & 1.3824 & 9.26 \\
         & \textit{FRB HTR} & 21:29:39.759, $-$79:28:32.50 & 0.000054 & 18.52 \\
         & \textit{Vela} & 08:35:20.65525, $-$45:10:35.1545 & 0.00268 & 9.26 \\
         & \textit{FRB RFI subtraction} & 21:29:39.70836, $-$79:28:32.2845 & 0.016 & 9.26 \\
         & \textit{FRB HTR RFI subtraction} & 21:29:39.759, $-$79:28:32.50 & 0.016 & 18.52 \\
         & \textit{Vela RFI subtraction} & 08:35:20.65525, $-$45:10:35.1545 & 0.00893 & 9.26 \\ \hline
        FRB~190608 & \textit{FRB calibrator} & 19:39:25.0263, $-$63:42:45.624 & 1.3824 & 9.26 \\
         & \textit{FRB position} & 22:16:07, $-$07:54:00 & 0.01 & 9.26 \\
         & \textit{FRB continuum field} & 22:15:26.3, $-$08:13:24 & 1.3824 & 9.26 \\
         & \textit{FRB HTR} & 22:16:04.75, $-$07:53:53.6 & 0.000216 & 9.26 \\
         & \textit{Vela} & 08:35:20.5193, $-45$:10:34.287 & 0.0036 & 9.26 \\
         & \textit{FRB RFI subtraction} & 22:16:07, $-$07:54:00 & 0.060 & 9.26 \\
         & \textit{FRB HTR RFI subtraction} & 22:16:04.75, $-$07:53:53.6 & 0.0235 & 9.26 \\
         & \textit{Vela RFI subtraction} & 08:35:20.5193, $-45$:10:34.287 & 0.014 & 9.26 \\ \hline
        FRB~190611 & \textit{FRB calibrator} & 04:08:20.380, $-$65:45:09.08 & 1.3824 & 9.26 \\
         & \textit{FRB position} & 21:23:00, $-$79:24:00 & 0.002 & 9.26 \\
         & \textit{FRB continuum field} & 21:23:00, $-$79:24:00 & 1.3824 & 9.26 \\
         & \textit{FRB HTR} & 21:22:59.11, $-$79:23:51.9 & 0.000108 & 9.26 \\
         & \textit{Vela} & 08:35:20.5193, $-45$:10:34.287 & 0.0036 & 9.26\\
         & \textit{FRB RFI subtraction} & 21:23:00, $-$79:24:00 & 0.031 & 9.26 \\
         & \textit{FRB HTR RFI subtraction} & 21:22:59.11, $-$79:23:51.9 & 0.031 & 9.26 \\
         & \textit{Vela RFI subtraction} & 08:35:20.5193, $-45$:10:34.287 & 0.014 & 9.26 \\ \hline
        FRB~190711 & \textit{FRB calibrator} & 04:08:20.380, $-$65:45:09.08 & 1.3824 & 9.26 \\
         & \textit{FRB position} & 21:57:40.012, $-$80:21:28.18 & 0.013176 & 9.26 \\
         & \textit{FRB continuum field} & 21:57:12.115, $-$80:26:3.025 & 1.3824 & 9.26 \\
         & \textit{FRB HTR} & 21:57:40.012, $-$80:21:28.18 & 0.000216 & 9.26 \\
         & \textit{Vela} & 08:35:20.65525, $-$45:10:35.1545 & 0.00357 & 9.26 \\
         & \textit{FRB HTR RFI subtraction} & 21:57:40.012, $-$80:21:28.18 & 0.032 & 9.26 \\
         & \textit{Vela RFI subtraction} & 08:35:20.65525, $-$45:10:35.1545 & 0.00715 & 9.26 \\ \hline
    \end{tabular}
    \caption{Parameters used in the correlation to produce the visibility datasets for each FRB in the sample}
    \label{tab:vis_datasets}
\end{table*}
\subsection{Determining FRB positions and uncertainties} \label{sec:FRBpos_det}
A full description of the process used to determine the final FRB positions and uncertainties is given in \cite{Bannister565}, \cite{Prochaska231}, and \cite{Macquart2020_DMz}. In brief, the \textit{FRB position} and \textit{FRB continuum} visibilities were imaged using the \textsc{casa} task \textsc{tclean} for each FRB in our sample after calibration, RFI subtraction, and optimally weighting the visibilities in frequency \citep{Bannister565,Prochaska231}, with the latter two only done for the \textit{FRB position} data. In the cases of FRB~190711 and FRB~190608, 
a time-independent frequency weighting did not result in an optimal signal-to-noise ratio (S/N). Accordingly, for these \textit{FRB position} datasets, we weighted the visibilities in time, as described in Section \ref{sec:htr_data_descript}, prior to the standard frequency weighting undertaken for the \textit{FRB position} datasets for all FRBs in our sample, following the method described in \cite{Bannister565} and \cite{Prochaska231}. The \textit{FRB continuum} and \textit{FRB position} visibilities were imaged in widefield, multi-scale multi-frequency synthesis\footnote{specmode and deconvolver were set to mfs and multiscale, respectively} mode with natural weighting and, for the former, one or two Taylor terms, depending on the field sources. The statistical position and uncertainty were obtained via the \textsc{aips} task \textsc{jmfit}, which fits a 2-D Gaussian to a region of an image. Here, the selected region of the total intensity \textit{FRB position} image was roughly the size of the synthesised beam and was centred on the FRB.

Given the phase solutions derived from the \textit{FRB calibrator} are extrapolated temporally and spatially when applied to the target datasets, the calibrated \textit{FRB position} data are subject to systematic positional offsets. However, since the \textit{FRB continuum} data contain the FRB signal and are calibrated with the same phase solutions, they are identically affected and can, therefore, be used to correct the FRB position and estimate the final positional uncertainty. To that end, the positions of any background radio sources detected in the total intensity \textit{FRB continuum} image were extracted using \textsc{jmfit} and compared to positions obtained from a reference image in order to tie the ASKAP frame to the ICRF3. For FRB~180924, FRB~190102, FRB~190611, and FRB~190711, data taken with the Australian Telescope Compact Array (ATCA), which has a comparable angular and frequency resolution -- thus reducing potential offsets in the fit centroids due to source structure -- was used to make the reference image. An image from the Faint Images of the Radio Sky at Twenty centimetres (FIRST) survey \citep{Becker95}, which has approximately twice the ASKAP angular resolution, was used as the reference for FRB~190608. As described in \cite{Macquart2020_DMz}, we assumed any calibration errors led to a simple translation of the FRB field and used the offsets in the background radio continuum sources to measure and correct this effect. As shown in Table~\ref{tab:offsets}, the offsets for FRB~180924 \citep{Bannister565} and FRB~190102 were consistent with zero, while the maximum offset (for FRB~190611) was 1.67 arcsec.
%
\begin{table}
    \centering
    \begin{threeparttable}
    \begin{tabular}{lll}
    \hline
        FRB & weighted mean offset & uncertainty\tnote{($\dagger$)} \\
         & (R.A., Decl. arcsec) & (R.A., Decl. arcsec) \\ \hline \hline
        180924\tnote{(*)} & 0.0, 0.0 & 0.09, 0.09 \\
        190102 & 0.0, 0.0 & 0.4, 0.5 \\
        190608 & 0.4, $-$0.9 & 0.2, 0.2 \\
        190611 & 1.7, 0.2 & 0.6, 0.6 \\
        190711 & 1.7, $-0.4$ & 0.4, 0.3 \\
    \hline
    \end{tabular}
    \begin{tablenotes}[flushleft]
        \item[($\dagger$)] For FRBs with offsets consistent with zero, the final systematic uncertainty listed here is the quadrature sum of the background source uncertainties, using the method described in \cite{Bannister565} and \cite{Prochaska231}.
        \item[(*)] The offset and uncertainty are from \cite{Bannister565}.
    \end{tablenotes}
    \end{threeparttable}
    \caption{The weighted mean offset and uncertainty values for the FRBs in our sample derived using (unless otherwise noted) the method described in \protect\cite{Macquart2020_DMz}.}
    \label{tab:offsets}
\end{table}
\subsection{Full polarisation imaging and flux density extraction} \label{sec:htr_data_descript}
For each FRB in our sample, after the RFI in the \textit{FRB HTR} visibility dataset was mitigated and the data calibrated as described in Section \ref{sec:method}, full polarisation imaging was performed for each integration timestep separately using the \textsc{casa}\footnote{All images discussed in this work were made with either CASA 5.3.0-143 or CASA 5.5.0-149} task \textsc{tclean}. The images were made using the \textsc{tclean} widefield, multi-scale cube mode with natural weighting for each visibility. Two imaging phase centres were used: one at the location of the FRB, as determined by the \textit{FRB position} dataset, and one offset by 5 arcminutes in right ascension and 5 arcminutes in declination to obtain an image rms estimate in a signal-free region. The frequency-averaged and dynamic spectra (Figures  \ref{fig:fscrunch} and \ref{fig:dynspec}) were then obtained by extracting the flux density (in units of jansky/beam) of the central pixel in each frequency-averaged slice of the image cube for all timesteps in the \textit{FRB HTR} dataset using the \textsc{imstat} task in \textsc{casa} to determine the maximum flux density value at the FRB position and, in the case of the former, subsequently averaging over frequency for each timestep. The rms was derived over a central region enclosing 75 percent of the noise estimation image via \textsc{imstat}.

For most of our sample, the statistical uncertainty of the FRB position was negligible in comparison to the uncertainty on the systematic shift in the reference frame estimated from the position of background sources. For FRB~190608 and FRB~190711, however, this was not the case. These wide FRBs did not gain as much from the high time resolution over the detection S/N, and both had relatively small uncertainties in the systematic shift estimation. Accordingly, to maximise our S/N and hence minimise the statistical position uncertainty in these cases, we used the \textit{FRB HTR} Stokes~I spectrum to temporally reweight the final \textit{FRB position} dataset used to obtain the FRB position and its statistical uncertainty. Unlike the other FRBs, which used a simple on/off gate for the \textit{FRB position} dataset, the FRB~190608 and FRB~190711 \textit{FRB position} data were correlated using the amplitudes obtained from their \textit{FRB HTR} frequency-averaged spectra as weights for each of the timesteps used to create the \textit{FRB HTR} visibilities if they exceeded a threshold of $\sim0.2$~Jy (FRB~190711) or $\sim0.8$~Jy (FRB~190608), where the threshold was dictated by the burst temporal structure (zero otherwise). These were averaged together to form a single weighted visibility. Compared to a simple on/off gate, this method results in a higher S/N and, therefore, improved statistical uncertainties. In our sample of 5 FRBs, however, FRB~190608 and FRB~190711 are the only ones for which the statistical uncertainty would have dominated the final positional uncertainty using a simple on/off gate, and hence the only ones that benefit significantly from this additional processing. As with the other \textit{FRB position} datasets, the FRB~190608 and FRB~190711 visibility datasets were optimally weighted by frequency following the method described in \cite{Bannister565}.
\subsection{Polarisation calibration} \label{sec:pol_cal}
In order to explore the polarisation properties of the FRBs in our sample, observations of the pulsar PSR J0834$-$4510 (the \textit{Vela} datasets described in Section \ref{sec:method}) were used to correct for instrumental polarisation leakage and determine both the rotation measure (RM) and absolute linear polarisation position angle (PA) of each burst.

When a burst propagates through a cold plasma containing an ordered magnetic field ($\vec{B}$), the component parallel to the line-of-sight (${B}_\parallel$) will induce generalised Faraday rotation in the polarisation direction of the linearly polarised light. The modified linear polarisation position angle (PA) can be modelled as
\begin{equation} \label{eq:pa}
    \Psi(\nu) = \Psi_0 + \rm RM c^2 (\nu^{-2} - \nu_0^{-2}) ,
\end{equation}
\noindent where $\Psi_0$ is the PA defined at a reference frequency $\nu_0$ (the centre of the band for each burst in our sample; see Figure \ref{fig:fscrunch}), and the rotation measure (RM) is defined as
\begin{equation} \label{eq:rm}
    \text{RM} \equiv \frac{e^3}{2 \pi m_e^2 c^4} \int_{d}^{0} \frac{B_\parallel(l) n_e(l)}{(1+z)^2} dl,
\end{equation}
\noindent where $e$ and $m_e$ are the electron charge and mass, respectively; $n_e$ is the electron density at $l$; and $d$ is the distance to the source. Here, we report the observed RM and do not correct it to the source reference frame. Since the linearly polarised Phased Array Feeds (PAFs) used in the ASKAP system can be rotated with respect to the nominal ordinal axes due to a third axis on which the dishes can rotate \citep{Hotan2014PASA...31...41H,McConnell2016PASA...33...42M}, they can likewise be rotated with respect to $\Psi$, and we use an angle $\Delta \Psi$ to model the unknown amount of resultant conversion between Stokes~Q and U that would be measured by a perfect receiving system:
\begin{align}
    Q'(\nu) & = Q {\rm cos}(\Delta \Psi) + U {\rm sin}(\Delta \Psi) \label{eq:q_deltaPsi} \\
    U'(\nu) & = -Q {\rm sin}(\Delta \Psi) + U {\rm cos}(\Delta \Psi) \label{eq:u_deltaPsi},
\end{align}
\noindent where $Q'(\nu)$ and $U'(\nu)$ are the rotated Stokes~Q and U; $U = L {\rm sin}(2\Psi(\nu))$ and $Q = L {\rm cos}(2\Psi(\nu))$ are the Faraday rotated Stokes parameters; and the total linear polarisation $L = \sqrt{Q^2 + U^2}$.

Finally, the ASKAP PAFs are linearly polarised: accordingly, instrumental delay and phase offsets between the two polarisations could lead to polarisation leakage. Here, we assume these offsets to be the sole source of this leakage, resulting in rotation between only Stokes~U and V. The observed Stokes parameters can then be described by
\begin{align}
    Q_{\rm obs}(\nu) & = Q'(\nu) \label{eq:q_obs}\\
    U_{\rm obs}(\nu) & = U' {\rm cos}(\Phi + 2\pi \nu \Delta t) + V {\rm sin}(\Phi + 2\pi \nu \Delta t) \label{eq:u_obs} \\
    V_{\rm obs}(\nu) & = -U' {\rm sin}(\Phi + 2\pi \nu \Delta t) + V {\rm cos}(\Phi + 2\pi \nu \Delta t) \label{eq:v_obs},
\end{align}
\noindent where $\Delta t$ and $\Phi$ are respectively the instrumental delay and phase offsets between the measured horizontal and vertical linear polarisations. To model the instrumental leakage, we compare ASKAP observations of Vela (\textit{Vela} datasets) to a well-calibrated observation in the same band observed with the 64-m Parkes radio telescope. We use nested sampling to measure $L$, $\Delta \Psi$, $\Phi$, and $\Delta t$ by fitting equations \ref{eq:q_obs} to \ref{eq:v_obs} to the \textit{Vela} data. Table~\ref{tab:vela_cal} shows the derived parameters.

Using the measured values of the Stokes parameters (Section \ref{sec:htr_data_descript}) in each frequency channel $i$, we apply a series of steps to calibrate each timestep of the data. First, we de-rotate $U_{\rm obs}$ and $V_{\rm obs}$ to correct for the instrumental leakage (i.e., swapping the signs of the sines in equations \ref{eq:u_obs} to \ref{eq:v_obs}). As recent tests of the ASKAP system have indicated that the PAF basis is left-handed, in order to follow the PSR/IEEE convention for the Stokes parameters \citep{van_straten_manchester_johnston_reynolds_2010}, the sign of Stokes~Q is then negated. Finally, we de-rotate $Q'(\nu)$ and $U'(\nu)$ to account for the unknown angle at which the PAFs are rotated relative to $\Psi$. The combined steps are applied via the following
\begin{align}
    Q_{i} = & -Q_{{\rm obs},i} {\rm cos} \Delta \Psi - [U_{{\rm obs},i} {\rm cos}(\Phi + 2\pi \nu_i \Delta t) \nonumber \\ & - V_{{\rm obs},i} {\rm sin}(\Phi + 2\pi \nu_i \Delta t)] {\rm sin} \Delta \Psi \label{eq:q_cal} \\
    U_{i} = & -Q_{{\rm obs},i} {\rm sin} \Delta \Psi + [U_{{\rm obs},i} {\rm cos}(\Phi + 2\pi \nu_i \Delta t) \nonumber \\ & - V_{{\rm obs},i} {\rm sin}(\Phi + 2\pi \nu_i \Delta t)] {\rm cos} \Delta \Psi \label{eq:u_cal} \\
    V_{i} = & \text{ } U_{{\rm obs},i} {\rm sin}(\Phi + 2\pi \nu_i \Delta t) + V_{{\rm obs},i}{\rm cos}(\Phi + 2\pi \nu_i \Delta t). \label{eq:v_cal}
\end{align}
Note that Vela was observed at the beam centre and any frequency dependence in the polarisation leakage due to the ASKAP PAF beam weights used in each observation is not accounted for in this procedure, so any small variations within the data are not captured. These are likely consistent with the observed leakage in FRB~181112 reported by \citet{Cho2020arXiv200212539C} -- i.e., $\lesssim2$\% at roughly the half power point -- as the FRBs in our sample are all within the half power point.
\subsection{Extracting derived parameters} \label{sec:derived_params}
\subsubsection{Rotation measures and polarisation position angles} \label{subsec:RMs_PAs}
After applying the derived calibration solutions, we search the corrected Stokes~Q and U for Faraday rotation using a modified version of the likelihood method described in \cite{Bannister565} and \cite{Prochaska231} and then use these to correct for the Faraday rotation in each FRB. We use the nested samples from the calibration solution for the parameters $\Delta \Psi$, $\Phi$, and $\Delta t$ to marginalise over uncertainty in the calibration
solution. We model the linear polarised flux to be
\begin{align}
    \hat{Q_i} = L_i {\rm cos}(2 {\rm RM} (\lambda_i^2 - \lambda_0^2) + 2\chi_0) \label{eq:q_hat} \\
    \hat{U_i} = L_i {\rm sin}(2 {\rm RM} (\lambda_i^2 - \lambda_0^2) + 2\chi_0) \label{eq:u_hat},
\end{align}
\noindent where $\chi_0$ is the PA at a reference wavelength $\lambda_0 = c/\nu_0$. We assume the noise is identical across frequency channels and between Stokes~Q and U when applying the maximum likelihood estimation. While this is not strictly the case, the differences are small, and, therefore, the results are unlikely to change significantly. For all FRBs, the PA was integrated over the entire pulse profile in order to determine their RMs. Additionally, for FRB~190102 and FRB~190611, the PA was integrated over each sub-burst region to calculate the RMs for the individual sub-bursts. Table~\ref{tab:vela_cal} shows the derived RMs. Once the RMs for each burst (or sub-burst) were determined, the calibrated data were de-rotated using the following
\begin{align}
    Q_{{\textrm{de-RM}},i} = Q_{i} {\rm{cos}}(2\psi_{\textrm{RM},i}) + U_{i} {\textrm{sin}}(2\psi_{\textrm{RM},i}) \label{eq:q_de-rm} \\
    U_{{\textrm{de-RM}},i} = U_{i} {\textrm{cos}}(2\psi_{\textrm{RM},i}) - Q_{i} {\textrm{sin}}(2\psi_{\textrm{RM},i}) \label{eq:u_de-rm}
\end{align}
\noindent where $\psi_{\textrm{RM},i} = {\rm RM} (\lambda_i^2 - \lambda_0^2)$.

The de-rotated spectra were then averaged over frequency (bottom panel of Figure \ref{fig:fscrunch}) and used to both remove the bias in the total linear polarisation, $L$, and determine the absolute PA for each FRB along with de-biasing it. The Faraday rotation corrected PA is given by
\begin{equation}
    \Psi_{{\textrm{de-RM}}} = \frac{1}{2} {\rm tan^{-1}} \left( \frac{U_{{\textrm{de-RM}}}}{Q_{{\textrm{de-RM}}}} \right). \label{eq:PA_deRMd}
\end{equation}
\noindent Following \cite{Everett&Weisberg2001}, we remove the bias in the derived $L$, $\Psi_{{\textrm{de-RM}}}$, and the uncertainty in $\Psi_{{\textrm{de-RM}}}$, where the latter is determined by propagation of uncertainties to be
\begin{equation}
    \sigma_\Psi^2 = \frac{Q_{{\textrm{de-RM}}}^2\sigma_U^2 + U_{{\textrm{de-RM}}}^2\sigma_Q^2}{4(Q_{{{\textrm{de-RM}}}}^2+U_{{\textrm{de-RM}}}^2)^2} , \label{eq:sigmaPA}
\end{equation}
\noindent where $\sigma_U$ and $\sigma_Q$ are the rms in Stokes~U and Q, respectively, obtained from the noise image (see Section \ref{sec:htr_data_descript}). Note that we compared the rms values and found $\sigma_U = \sigma_Q = \sigma_I$ to within 1\%, satisfying this assumption in \cite{Everett&Weisberg2001}.

The frequency-averaged, de-biased total linear polarisation, $L_{{\textrm{de-bias}}}$, is calculated using Equation 11 in \cite{Everett&Weisberg2001} (corrected here for a typographical error):
\begin{equation} \label{eq:L_de-bias}
    L_{{\textrm{de-bias}}} =
    \begin{cases}
      \sigma_I \sqrt{\left(\frac{L_{{\rm meas}}}{\sigma_I}\right)^2 - 1} & \text{if $\frac{L_{{\rm meas}}}{\sigma_I} > 1.57$} \\
      0 & \text{otherwise} .
    \end{cases}
\end{equation}
\noindent Using $L_{{\textrm{de-bias}}}$ and a user-defined threshold of $2\sigma_I$, we then mask $\Psi_{\textrm{de-RM}}$ and $\sigma_\Psi$ values where the following conditions are true: $L_{{\textrm{de-bias}}} < 2\sigma_I$ and $L_{{\textrm{de-bias}}} = 0$. These correspond to low S/N data points, and their removal effectively de-biases $\Psi_{\textrm{de-RM}}$ and $\sigma_\Psi$, as the high S/N values are less affected by these biases. For $\Psi_{{\textrm{de-RM}}}$ values with a mean near $\pm 90 \degree$, as was the case for FRB~190711, we also correct for phase wrapping by adding $180\degree$ to values less than zero. The non-masked values of $\Psi_{{\textrm{de-RM}}}$ are plotted in the top panels of Figure \ref{fig:fscrunch}, where the error bars are the non-masked values of $\sigma_\Psi$.
\begin{figure*}
\centering
\captionsetup[subfigure]{labelformat=empty}
\subfloat[]{\includegraphics[width=0.452\linewidth]{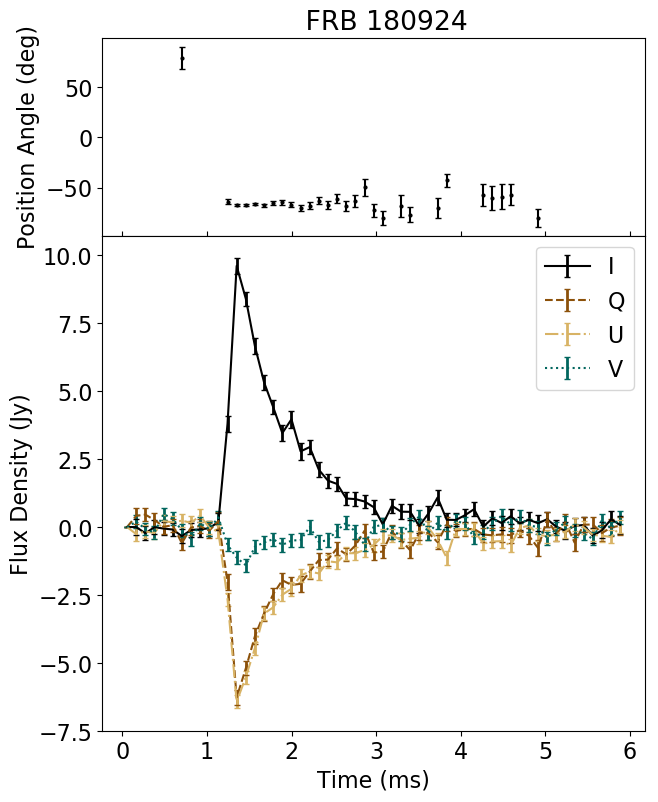}}
\subfloat[]{\includegraphics[width=0.43\linewidth]{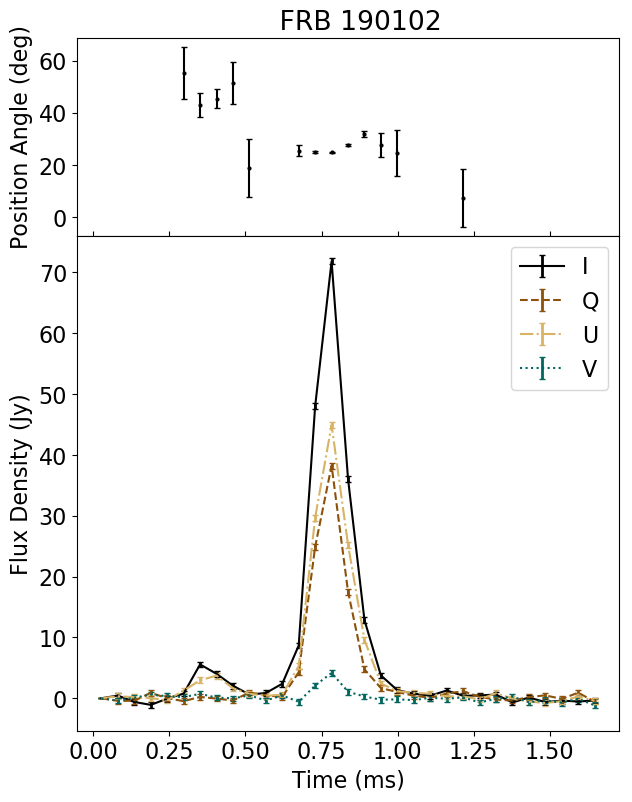}}\\[-6ex]
\subfloat[]{\includegraphics[width=0.442\linewidth]{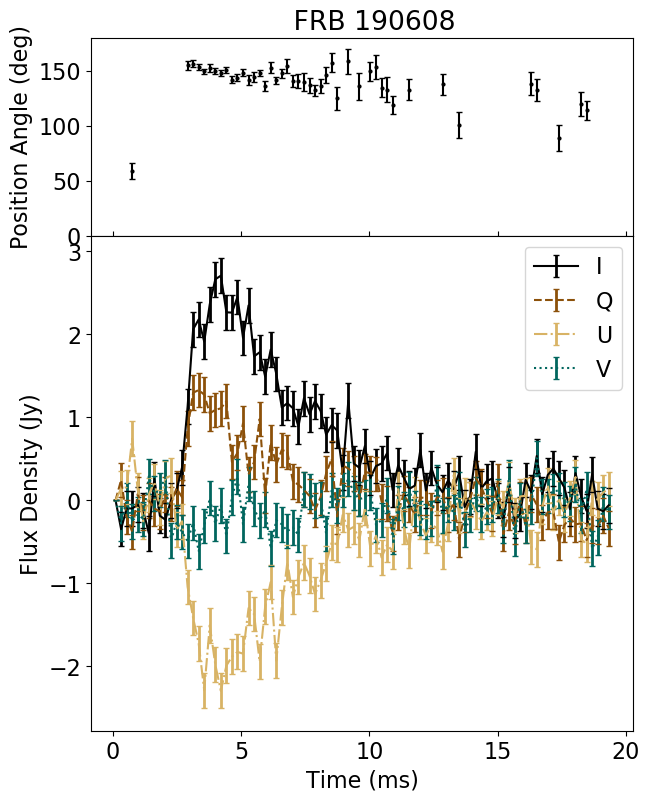}}
\subfloat[]{\includegraphics[width=0.43\linewidth]{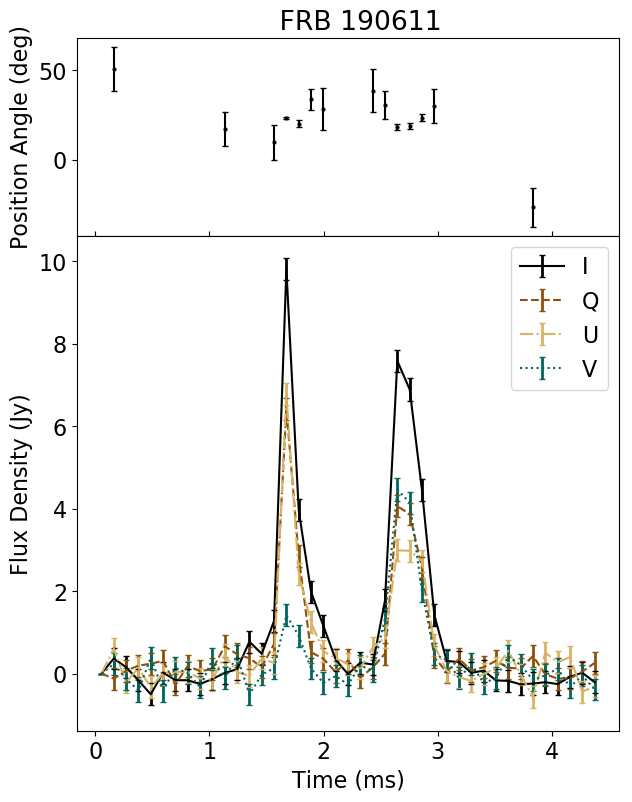}}\\[-5ex]
\caption{Spectropolarimetric properties of our sample of FRBs. Top panels: polarisation position angle versus time, referenced to the centre of the band. Bottom panels: frequency averaged time series. Reading left to right and then top to bottom: FRB~180924, FRB~190102, FRB~190608, FRB~190611.
\label{fig:fscrunch}}
\end{figure*}

\begin{figure*}
    \centering
    \captionsetup[subfigure]{labelformat=empty}
    \subfloat[]{\includegraphics[width=0.45\linewidth]{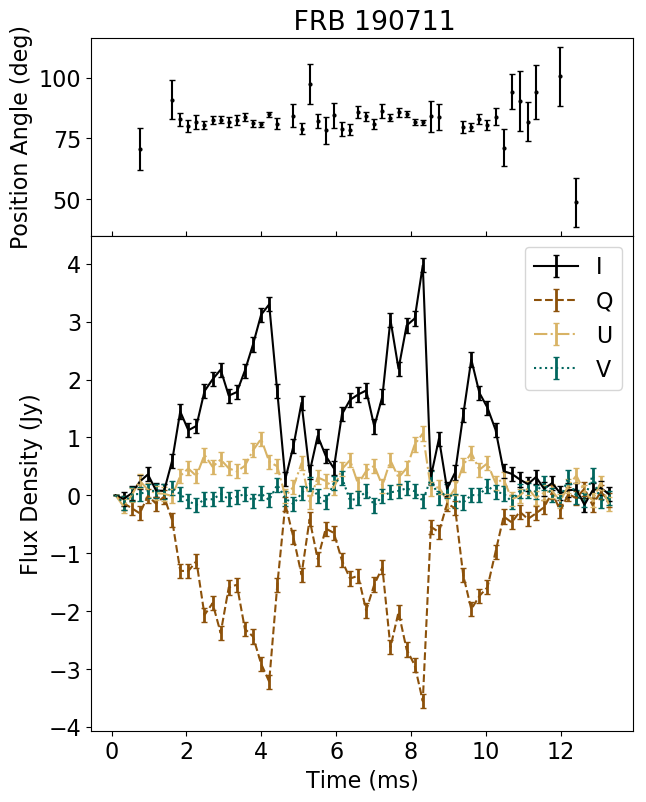}}\\[-4.5ex]
    \contcaption{Same caption as the above. Shown here are the results for FRB~190711.}
    \label{fig:fscrunch_continued}
\end{figure*}

\begin{figure*}
    \centering
    \captionsetup[subfigure]{labelformat=empty}
    \subfloat[]{\includegraphics[width=0.2\textwidth]{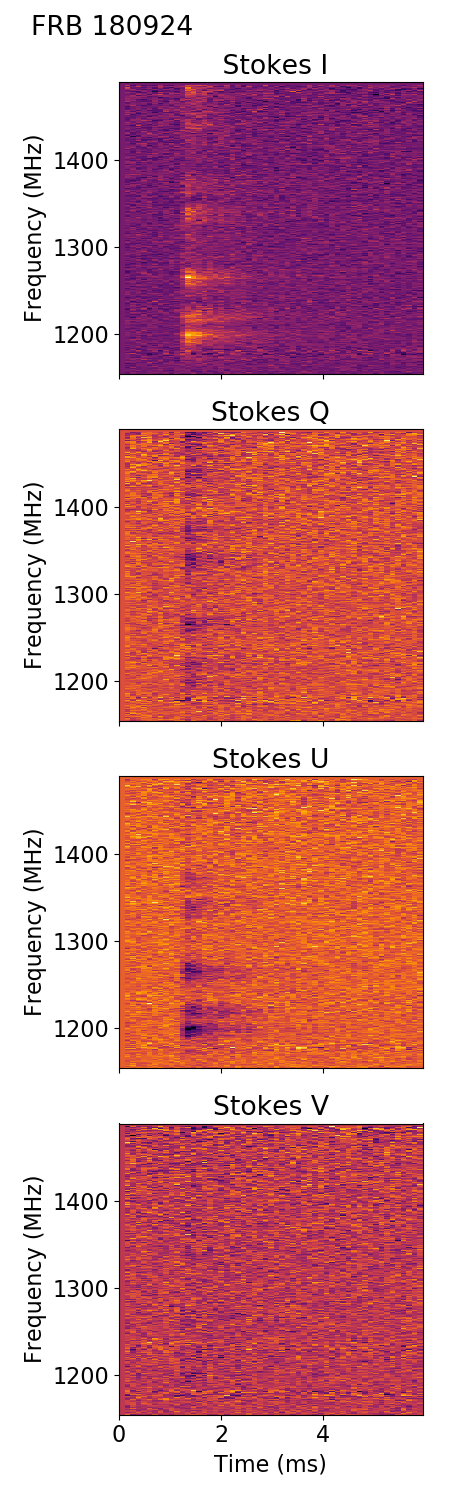}}
    \subfloat[]{\includegraphics[width=0.2\textwidth]{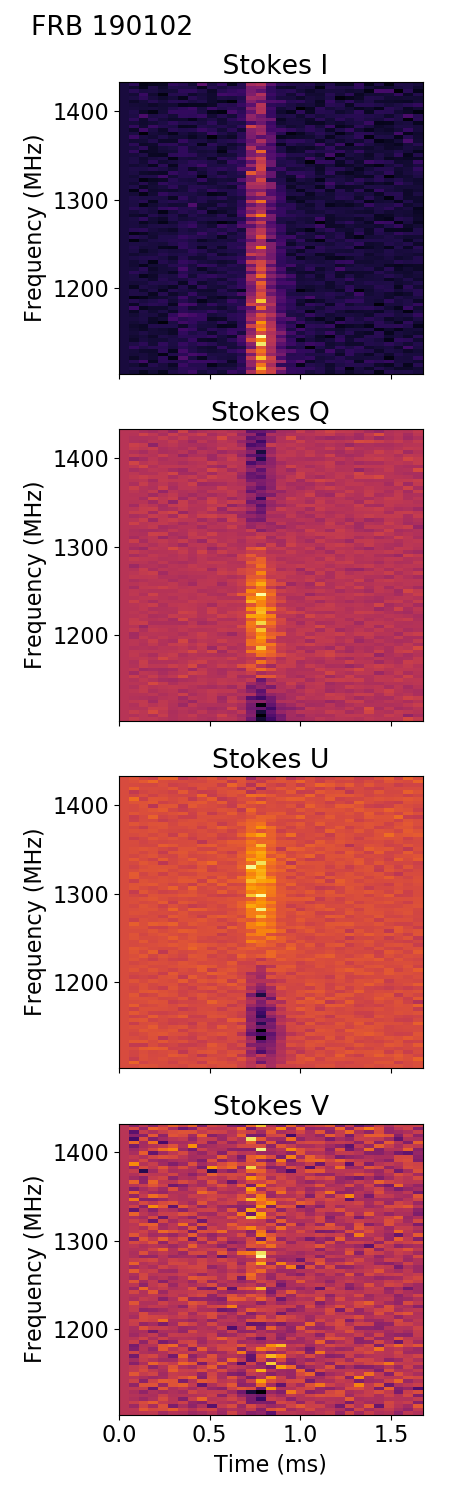}}
    \subfloat[]{\includegraphics[width=0.2\textwidth]{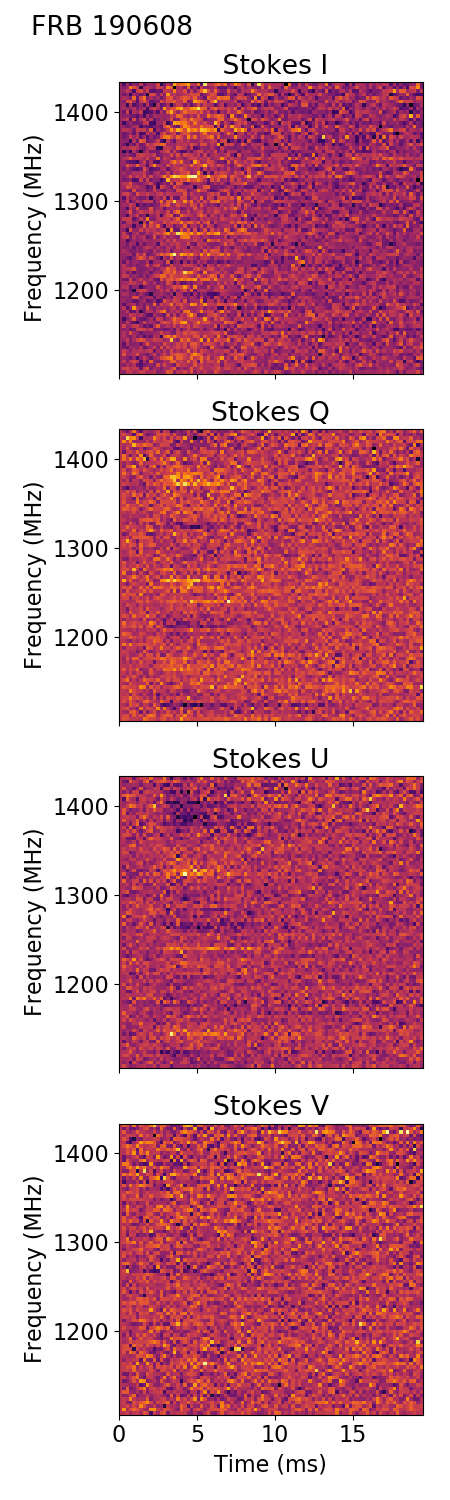}}
    \subfloat[]{\includegraphics[width=0.2\textwidth]{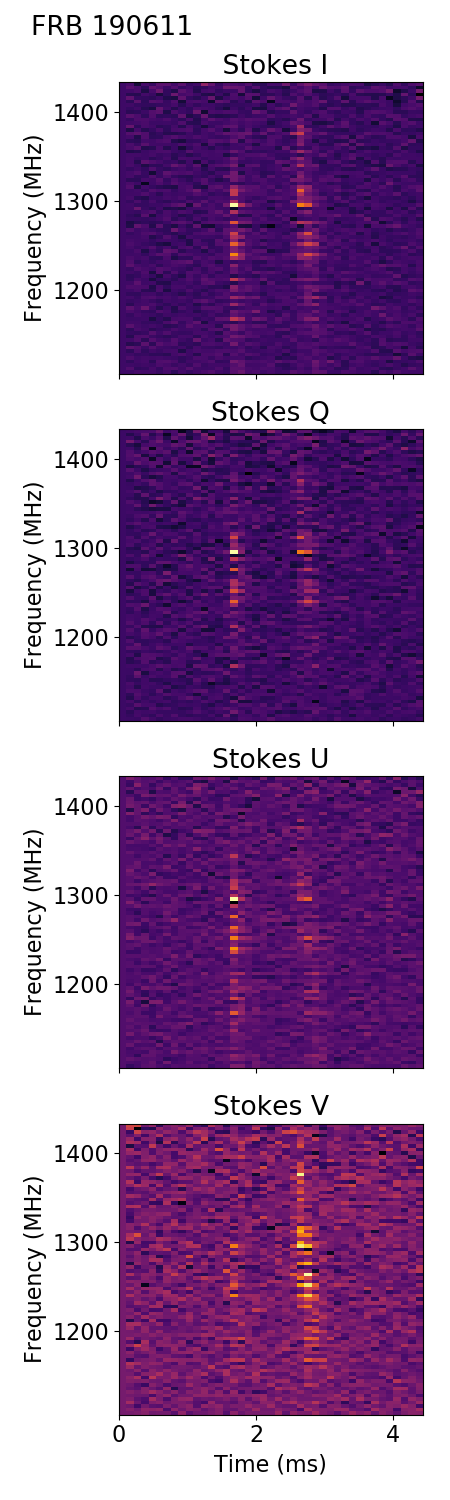}}
    \subfloat[]{\includegraphics[width=0.2\textwidth]{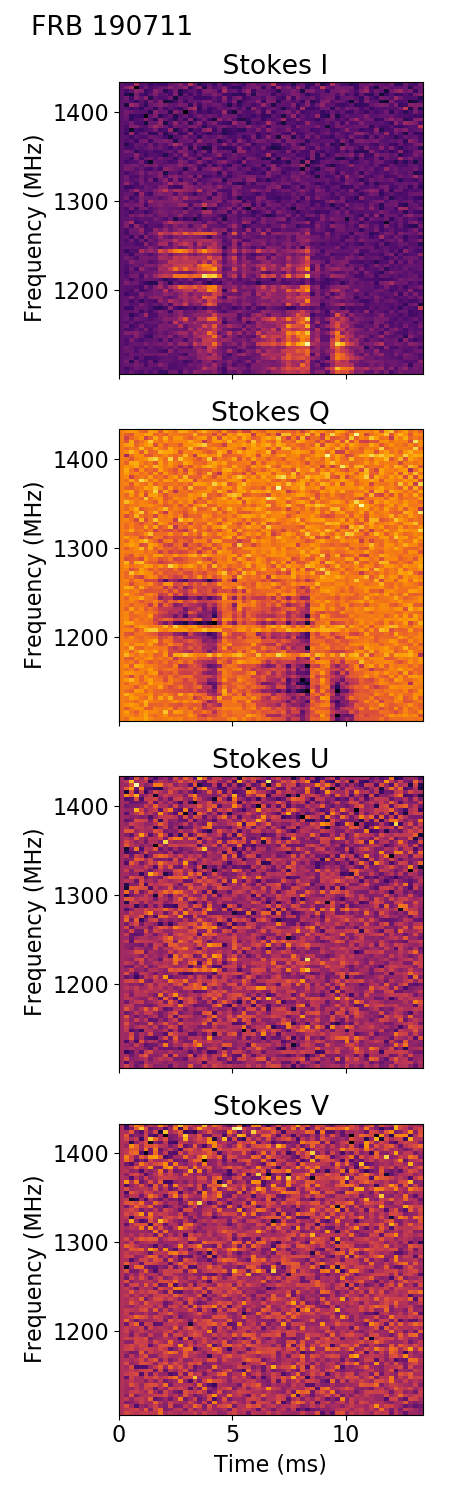}}
    \caption{Dynamic spectra for the sample of FRBs. Reading left to right: FRB~180924, FRB~190102, FRB~190608, FRB~190611, FRB~190711. The colour corresponds to the flux density, with each subplot auto-scaled such that white and black respectively correspond to the most positive and most negative values in the sub-plot.}
    \label{fig:dynspec}
\end{figure*}

\subsubsection{Polarisation fractions}
We use the calibrated Stokes parameters to derive polarisation fractions for each FRB in the sample. The total intensity, $I$, and its uncertainty, $\sigma_I$, are given by the measured, frequency-averaged Stokes~I flux density and rms (the latter from the Stokes~I noise image; see Section \ref{sec:htr_data_descript}), respectively. Similarly, the total circular polarisation, $V$, and its uncertainty, $\sigma_V$, are derived from the calibrated Stokes~V flux density (Equation \ref{eq:v_cal}) and noise image rms, averaged over frequency. The total linear polarisation is given by Equation \ref{eq:L_de-bias} and its uncertainty by
\begin{equation}
    \sigma_L^2 = \frac{Q_{{{\textrm{de-RM}}}}^2 \sigma_Q^2 + U_{{{\textrm{de-RM}}}}^2 \sigma_U^2}{L_{\textrm{de-bias}}^2}. \label{eq:sigma_L}
\end{equation}
\noindent The total polarisation and its uncertainty are determined via
\begin{align}
    P & = \sqrt{L_{\textrm{de-bias}}^2 + V^2} \label{eq:P} \\
    \sigma_P^2 & = \frac{Q_{{{\textrm{de-RM}}}}^2 \sigma_Q^2 + U_{{{\textrm{de-RM}}}}^2 \sigma_U^2 + V^2 \sigma_V^2}{L_{\textrm{de-bias}}^2 + V^2}, \label{eq:sigma_P}
\end{align}
where we note that the lack of de-biasing in Stokes~V would only affect calculations of $P$ when the total polarisation is low, which is not the case for any of our FRBs.

These can then be combined to determine the total weighted average polarisation fractions (i.e., relative to the total intensity) for each burst or sub-burst within an FRB. In order to calculate these, we first determine each the weighted average (i.e., $I_{\rm was}$, $P_{\rm was}$, $L_{\rm was}$, and $V_{\rm was}$) and weighted average noise (i.e., $\sigma_{I,\rm wan}$, $\sigma_{P,\rm wan}$, $\sigma_{L,\rm wan}$, and $\sigma_{V,\rm wan}$) over time ranges corresponding to individual bursts within the total signal envelope. With $\rho = \{I, P, L, V\}$, this results in the following
\begin{equation}
    \rho_{\rm was} = \frac{\sum\limits_{t=i}^n \rho(t) I(t)}{\sum\limits_{t=i}^n I(t)} \quad \pm \quad \sigma_{\rho,\rm wan} = \frac{\sqrt{\sum\limits_{t=i}^n \sigma_{\rho}^2(t) I^2(t)}}{\sum\limits_{t=i}^n I(t)}. \label{eq:Stokes_wan}
\end{equation}
\noindent We then take the ratios of these values relative to $I_{\rm was}$, with the uncertainties in these polarisation fractions given by
\begin{equation}
    \sigma_{\rho/I} = \frac{\sqrt{\sum\limits_{t=i}^n \sigma_{\rho,{\rm wan}}^2(t) + \frac{\rho_{\rm was}^2(t)}{I_{{\rm was}}^2(t)} \sigma_{I,{\rm wan}}^2(t)}}{\sum\limits_{t=i}^n I_{{\rm was}}(t)}. \label{eq:pol_frac_sigma}
\end{equation}
\noindent The polarisation fractions, their uncertainties, and the ranges of time over which the weighted sum were taken are listed in Table~\ref{tab:pol_fracs}.
\subsubsection{Differential dispersion measure: FRB~190611} \label{sec:deltaDM}

As seen in Figure~\ref{fig:dynspec}, the second sub-pulse for FRB~190611 exhibits a residual frequency-dependant arrival time delay after de-dispersion to a DM consistent with the best-fitting DM value from the first sub-pulse. Due to the patchy emission structure, it is not immediately apparent whether this frequency-dependent delay is consistent with a $\nu^{-2}$ dependence that would be expected for a differential dispersion measure, or if a different frequency dependence (which might indicate a different intrinsic origin) is preferred. 

To determine limits on the frequency dependence of the arrival time delay, we assumed the differential delay $d = A \nu^{\gamma}$, and performed a brute force search over the range $-4 < \gamma < 0$ and $0 < A < 3$ ms, where $\nu$ was expressed in GHz. Twenty-one grid points were used for both $\gamma$ and $A$.  The first half of the dynamic spectrum was excised to remove the first sub-pulse, and a first-order interpolation between adjacent data points in time was used to account for sub-sample shifts. For each trial, after each frequency channel was corrected, the resultant corrected dynamic spectrum was summed in frequency and the peak recorded.

\subsubsection{Scattering analysis} \label{sec:scatter_method}

\citet{qiu20} present a Bayesian framework to model the dynamic spectra of ASKAP FRBs to determine the maximum a-posteriori intrinsic width (assuming the intrinsic pulse morphology can be well described by a Gaussian component) and test for the presence of scattering caused by multipath propagation in an ionised medium. The results presented in \citet{qiu20} used the low time resolution data produced by the ASKAP search pipeline, but the methodology is applicable to our high time resolution data. We applied this same approach to the FRBs presented here, fitting only the Stokes~I polarisation and initially using one Gaussian component per FRB -- except for FRB~190611 where we use one component per sub-pulse. We did not attempt to model FRB~190711, which cannot be usefully represented by Gaussian components. We compare the Bayesian evidence between models ($\Delta \mathrm{LogE}$) with and without scattering to determine the favoured model. We report the 68\% credible intervals for intrinsic pulse width ($\sigma$), the best fit DM, scatter broadening time ($\rm \tau$), and frequency dependence of the scattering ($\alpha$) from the posterior distributions of the favoured model.

We further use this Bayesian framework to test two- and three-component models for FRB~180924 and a two-component model for FRB~190608 in order to determine if there is sufficient evidence for secondary components that are partially obscured by the scattering tails in these FRBs. Subsequent scattered Gaussian components are added to the model to account for any obscured component contributing to excess emission in the scattering tail.
\section{Results} \label{sec:results}
The FRBs in our sample are resolved in time and exhibit a wide variety of temporal and spectral morphologies as well as a range of RMs and polarisation properties, as can be seen in Figures \ref{fig:fscrunch} and \ref{fig:dynspec} and Tables \ref{tab:vela_cal} and \ref{tab:pol_fracs}. Figure \ref{fig:fscrunch} shows the full polarisation, high time resolution, frequency-averaged time series (flux density vs. time) for each FRB, while the dynamic spectra (frequency vs. time) for each Stokes parameter are shown in Figure \ref{fig:dynspec}. Table~\ref{tab:frb_props} lists the properties of each FRB, including the DM used in the production of the dynamic spectra and frequency-averaged plots, and it provides the best estimates for intrinsic pulse width, final dispersion measure, and scattering time for each FRB. Finally, Tables \ref{tab:vela_cal} and \ref{tab:pol_fracs} provide the derived RM values and pulse-averaged polarisation fractions, respectively.

As can be seen in Figure \ref{fig:fscrunch}, the pulse profiles exhibit a range of temporal and spectral features. All of the sources (with the exception of FRB~190711, where we did not attempt a scattering fit) show evidence for scattering with a frequency dependence similar to pulsar scattering caused by the ISM \cite[][]{Rickett1990}, with FRB~190102 having the narrowest scattering tail ($0.041^{+0.002}_{-0.003}$~ms) and FRB~190608 having the longest ($3.3 \pm 0.2$~ms). Three of the five FRBs display obvious temporal structure in addition to a scattering tail, with FRB~190102 and FRB~190611 having two sub-pulses and FRB~190711 having three distinct sub-bursts within its burst envelope. Following \cite{Hessels2019ApJ...876L..23H}, we define a sub-burst as being a clearly distinguishable (by eye) component in time and frequency. The precise isolation of components is complicated by the burst morphology as well as scattering and will be further discussed in Section \ref{sec:190711}. Substructure for FRB~180924 and FRB~190608 is obscured by the scattering and discussed in Section~\ref{sec:180924and190608}.

The dynamic spectra (Figure \ref{fig:dynspec}) reveal a range of spectral structure as well. FRB~190102 is relatively smooth across the band, while FRB~180924, FRB~190608, and FRB~190611 exhibit frequency banding of varying widths and the time-frequency structure of FRB~190711 is highly complex.

The polarisation properties also vary widely across the burst sample. The RM magnitudes range from $\rm 9 \pm 2 \,rad\,m^{-2}$ for FRB~190711 to $\rm 353 \pm 2 \,rad\,m^{-2}$ for FRB~190608 (Table~\ref{tab:vela_cal}), with the majority of FRBs having relatively low RMs. Of the FRBs with multiple components, the two sub-pulses within FRB~190102 and FRB~190611 have differential RMs (although in the case of FRB~190611, the difference is marginal), whereas the FRB~190711 burst envelope has a constant RM across all sub-bursts, within the measurement uncertainty. The behaviour of the PAs as a function of pulse phase also varies across the burst sample. While FRB~180924 and FRB~190711 have relatively flat PAs, FRB~190608 has a small but significant downward trend in PA across the burst profile. FRB~190102 and FRB~190611, in contrast, show evidence of PA swings within each of their sub-pulses. The pulse-averaged polarisation fractions seen in Table~\ref{tab:pol_fracs} also highlight the varied polarisation properties within the sample. FRB~180924 and FRB~190608 are highly linearly polarised with a non-negligible circular polarisation fraction, while FRB~190711 is consistent with being $100\%$ linearly polarised across its three sub-bursts. Conversely, the polarisation fractions evolve within and between the sub-pulses of both FRB~190102 and FRB~190611. While each sub-pulse in FRB~190102 remains highly linearly polarised with a non-zero component of circular polarisation, the total polarisation fraction increases between pulse 1 and 2. In contrast, the total polarisation fraction of FRB~190611 is consistent with remaining constant across the sub-pulses. However, the ratio of linear to circular polarisation changes significantly, with the second sub-pulse having a substantial circular polarisation fraction relative to its linear polarisation fraction.

As described in Section \ref{sec:deltaDM}, the second sub-pulse of FRB~190611 has a residual frequency-dependent delay in its arrival times when de-dispersed at the optimal DM for first sub-pulse. We find best-fitting values of $A=2.4$\,ms and $\gamma=-0.6$, but we are unable to significantly constrain $\gamma$, with values in the range $-2.6 < \gamma < -0.4$ all providing a peak flux density after correction within 1\,$\sigma$ of the best value ($A$ is of course highly covariant with $\gamma$, with values ranging from 0.8 to 3 ms).  The frequency-dependent delay seen in the second sub-pulse of FRB~190611 is therefore plausibly explained by a differential dispersion measure, but other origins cannot be excluded.

As discussed in Section \ref{sec:htr_data_descript}, the positions for FRB~190608 and FRB~190711 were improved by optimally weighting not only by frequency but also by time. Here, we update the positions and uncertainties given in \cite{Macquart2020_DMz}. While optimal weighting was used for FRB~190711, RFI subtraction for the \textit{FRB position} dataset was not previously used prior to reweighting in time and frequency. After applying RFI subtraction, its updated position and uncertainties are RA, Dec (J2000) =  21h57m40.68s $\pm \, 0.16$ (statistical; systematic: $\pm \, 0.048$; $\pm \, 0.15$), $-$80d21m28.8s $\pm \, 0.3$ (statistical; systematic: $\pm \, 0.07$; $\pm \, 0.3$). We note that while the statistical uncertainties have improved, the final position and astrometric uncertainties are unchanged from the \cite{Macquart2020_DMz} values, as these were already dominated by the systematic uncertainties as a result of the optimal weighting, and RFI subtraction does not improve the \textit{FRB continuum field} data. The FRB~190608 position and statistical uncertainty, which were derived from a non-optimally weighted \textit{FRB position} dataset for \cite{Macquart2020_DMz}, are also updated here. The final position and uncertainties are RA, Dec (J2000) = 22h16m4.77s $\pm \, 0.02$ (statistical; systematic: $\pm \, 0.01$; $\pm \, 0.01$), $-$07d53m53.7s $\pm \, 0.3$ (statistical; systematic: $\pm \, 0.2$; $\pm \, 0.2$). Of note, the median statistical precision in the positions of the FRBs in our sample is $\sim$0.1~arcsec in RA and $\sim$0.2~arcsec in Dec. Thus, if the systematic uncertainties could be reduced through improved calibration (for instance, if transfer of higher S/N calibration solutions from commensal ASKAP imaging observations can be commissioned), we would routinely get localisations at the $\sim$~0.1~-~0.2~arcsec level.
\begin{table*}
    \centering
    \begin{threeparttable}
    \begin{tabular}{lllllll}
        \hline
        FRB & $\Delta \Psi$ & $\Delta \rm t$ & $\Phi$ & $\rm RM_{Vela}$ & $\rm RM_{FRB}$ & $\rm RM_{MW}$ \tnote{$\dagger$}\\
         & (rad) & (ns) & (rad) & (rad $\rm m^{-2}$) & (rad $\rm m^{-2}$) & (rad $\rm m^{-2}$) \\
         \hline
         \hline
        FRB~180924 & $4.36 \pm 0.01$ & $-0.05 \pm 0.03$ & $-0.6 \pm 0.2 $ & $38.6 \pm 0.6$ & $22 \pm 2$ & $7 \pm 9$ \\
        FRB~190102 & $2.834 \pm 0.003$ & $-0.03 \pm 0.01$ & $0.3 \pm 0.1 $ & $42.8 \pm 0.2$ & $-105 \pm 1$ & $34 \pm 22 $ \\
        \multicolumn{1}{r}{pulse 1} & & & & & $-128 \pm 7$ \\
        \multicolumn{1}{r}{pulse 2} & & & & & $-105 \pm 1$ \\
        FRB~190608 & $2.923 \pm 0.004$ & $-0.06 \pm 0.02$ & $0.4 \pm 0.2 $ & $42.3 \pm 0.1$ & $353 \pm 2$ & $-25 \pm 8$ \\
        FRB~190611 & $2.961 \pm 0.008$ & $0.01 \pm 0.04$ & $-0.0 \pm 0.3 $ & $43.6 \pm 0.4$ & $20 \pm 4$ & $30 \pm 19$ \\
        \multicolumn{1}{r}{pulse 1} & & & & & $19 \pm 4$ \\
        \multicolumn{1}{r}{pulse 2} & & & & & $12 \pm 6$ \\
        FRB~190711 & $2.872 \pm 0.002$ & $0.10 \pm 0.01$ & $-0.82 \pm 0.09 $ & $43.7 \pm 0.1$ & $9 \pm 2$ & $27 \pm 20$ \\
        \multicolumn{1}{r}{sub-burst 1} & & & & & $10 \pm 2$ \\
        \multicolumn{1}{r}{sub-burst 2} & & & & & $9 \pm 3$ \\
        \multicolumn{1}{r}{sub-burst 3} & & & & & $12 \pm 6$ \\
        \hline
    \end{tabular}
    \begin{tablenotes}[flushleft]
        \item[$\dagger$] The estimates for the expected Galactic RM contribution are from \cite{Oppermann2015A&A...575A.118O} and were obtained via  \url{https://github.com/FRBs/FRB/blob/master/frb/rm.py}
    \end{tablenotes}
    \end{threeparttable}
    \caption{Maximum likelihood calibration parameters derived from Vela observations. $\rm RM_{Vela}$ and $\rm RM_{FRB}$ are the resultant RMs for Vela and the FRB, respectively, derived using the calibration solutions.}
    \label{tab:vela_cal}
\end{table*}
\begin{table*}
    \begin{center}
    \begin{threeparttable}
    \begin{tabular}{llllll}
    \hline \hline
        Source & FRB~180924 & FRB~190102 & FRB~190608 & FRB~190611\tnote{6} & FRB~190711\tnote{7} \\[2pt]
        $\rm t_{obs,FRB}$ (UTC) \tnote{(1)} & 16:23:12.562 & 05:38:44.002 & 22:48:13.370 & 05:45:43.421 & 01:53:41.690 \\[2pt]
        Number of antennas & 24 & 23 & 25 & 25 & 28 \\[2pt]
        Max. baseline ($\rm m$) & 5376 & 3946 & 5987 & 3975 & 4336 \\ [2pt]
        Correlation DM (pc cm$^{-3}$)\tnote{(2)} & 362.2 & 364.538 & 339.79 & 332.60 &  587.8683 \\[2pt]
        Calibrator & PKS 0407$-$658 & PKS 1934$-$638 & PKS 1934$-$638 & PKS 0407$-$658 & PKS 0407$-$658 \\[2pt]
        $\rm t_{obs,Cal}$ (UTC) \tnote{(3)} & 21:50:37.657 & 06:29:45.277 & 23:13:42.809 & 06:07:51.071 & 02:14:55.854 \\[2pt]
        RA (J2000, hh:mm:ss.s) & 21:44:25.255 $\pm$ 0.008 & 21:29:39.76 $\pm$ 0.17 & 22:16:04.77 $\pm$ 0.02 & 21:22:58.91 $\pm$ 0.25 & 21:57:40.68 $\pm$ 0.16 \\[2pt]
        Dec (J2000, dd:mm:ss.s) & $-$40:54:00.10 $\pm$ 0.11 &  $-$79:28:32.5 $\pm$ 0.5 & $-$07:53:53.7 $\pm$ 0.3 & $-$79:23:51.3 $\pm$ 0.7 & $-$80:21:28.8 $\pm$ 0.3 \\[2pt]
        $\ell$, $b$ (deg) & 0.742467, $-$49.414787 & 312.6537, $-33.4931$ & 53.2088, $-48.5296$ & 312.9352,  $-33.2818$ & 310.9078, $-33.9023$ \\[2pt]
         DM (pc cm$^{-3}$)\tnote{(4)} & 362.16 $\pm$ 0.01 & 364.545 $\pm$ 0.004 & 340.05$^{+0.06}_{-0.03}$ & 332.63 $\pm$ 0.04 &   \\[2pt]
        Pulse width $\sigma$ (ms) &$0.09\pm0.04$ & $0.053\pm0.002$ & $1.1\pm0.2$ & $0.09\pm0.02$ & \\[2pt]
        Scattering time $\rm \tau$ (ms) \tnote{(5)} & $0.68\pm 0.03$ & $0.041^{+0.002}_{-0.003}$ & $3.3\pm0.2$ & $0.18 \pm 0.02$ & \\[2pt]
        Scattering index $\alpha$ & $-3.6^{+0.6}_{-0.5}$ & $-3.84^{+0.71}_{-0.78}$ & $-3.5\pm0.9$ & $-5.86^{+1.73}_{-1.98}$ & \\[2pt]
        Bayesian Evidence $\Delta \mathrm{LogE}$ \tnote{(8)} &162 &17 &52 &11 & \\
    \hline \hline
    \end{tabular}
    \begin{tablenotes}[flushleft]
        \item[(1)] The time of the FRB observation; the UTC calendar day is given by the FRB name in YYMMDD format
        \item[(2)] Initial DM estimate used for the high time resolution correlation
        \item[(3)] The time of the calibrator observation; the calibrator scan was taken on the same UTC calendar day as the FRB
        \item[(4)] The final fit DM from the analysis described in Section \ref{sec:scatter_method}
        \item[(5)] Defined at a reference frequency of 1.2725 GHz
        \item[(6)] DM and scattering are reported for the first of the two sub-pulses for FRB~190611; differences between the two pulses are covered in the discussion
        \item[(7)] No attempt was made to fit the complex time-domain structure of FRB~190711, and so the final five rows are left intentionally blank for this FRB
        \item[(8)] The values listed here correspond to the evidence for scattering versus non-scattering models, where a positive value indicates that scattering is favoured
    \end{tablenotes}
    \end{threeparttable}
    \end{center}
    \caption{Properties of the sample of FRBs. The uncertainties on the RA and Dec are obtained by combining the statistical and systematic uncertainties in quadrature.}
    \label{tab:frb_props}
\end{table*}

\begin{table*}
    \centering
    \begin{tabular}{llllll}
        \hline
        FRB & $\frac{P_{\rm was}}{I_{\rm was}} \pm \sigma_{P_{\rm was}/I_{\rm was}}$ & $L_{\rm was}/I_{\rm was} \pm \sigma_{L_{\rm was}/I_{\rm was}}$ & $V_{\rm was}/I_{\rm was} \pm \sigma_{V_{\rm was}/I_{\rm was}}$ & $t_{\rm int}$ (ms) \\
         \hline
         \hline
        FRB~180924 & $91.3 \pm 2.0$ & $90.2 \pm 2.0$ & $-13.3 \pm 1.4$ & 1.08 - 3.24 \\
        FRB~190102 & & & & \\
        \multicolumn{1}{r}{pulse 1} & $70 \pm 8$ & $69 \pm 8$ & $9 \pm 7$ & 0.216 - 0.54 \\
        \multicolumn{1}{r}{pulse 2} & $82.3 \pm 0.7$ & $82.2 \pm 0.7$ & $4.8 \pm 0.5$ & 0.54 - 1.026 \\
        FRB~190608 & $92 \pm 3$ & $91 \pm 3$ & $-9 \pm 2 $ & 1.944 - 12.744  \\
        FRB~190611 & & & &  \\
        \multicolumn{1}{r}{pulse 1} & $94 \pm 3$ & $93 \pm 3$ & $15 \pm 2$ & 1.296 - 1.944 \\
        \multicolumn{1}{r}{pulse 2} & $91 \pm 3$ & $70 \pm 3$ & $57 \pm 3$ & 2.268 - 3.024 \\
        FRB~190711 & & & & \\
        \multicolumn{1}{r}{pulse 1} & $101 \pm 2$ & $101 \pm 2$ & $-1 \pm 2 $ & 0.216 - 4.536 \\
        \multicolumn{1}{r}{pulse 2} & $93.9 \pm 2.0$ & $93.7 \pm 2.0$ & $0.9 \pm 1.5$ & 4.536 - 8.856 \\
        \multicolumn{1}{r}{pulse 3} & $98 \pm 4$ & $98 \pm 4$ & $1 \pm 3$ & 8.856 - 11.448 \\
        \hline
    \end{tabular}
    \caption{The polarisation fractions along with their uncertainties derived for each FRB over the time range $t_{\rm int}$.}
    \label{tab:pol_fracs}
\end{table*}
\section{Discussion} \label{sec:discussion}
Where the FRB scattering is negligible compared to the intrinsic pulse width (FRB~190102, FRB~190611, and FRB~190711), Figure~\ref{fig:fscrunch} highlights the clear dichotomy between the broad and complex temporal structure (but simple polarimetric structure) of FRB~190711 and the narrow pulses with time-varying polarisation properties seen in FRB~190102 and FRB~190611.  For the two remaining FRBs, scattering obscures the underlying temporal and polarimetric structure, and the degree of similarity to these two categories is not immediately clear. Here, we consider each of these categories in turn.
\subsection{FRB~190711: footprints of a repeating FRB} \label{sec:190711}
The FRB~190711 burst exhibits many of the hallmarks of repeating FRBs. As with FRB~121102 \citep[e.g., ][]{Michilli2018Natur.553..182M}, FRB~190711 has a pulse-averaged linear polarisation fraction of approximately $100\%$ (Table~\ref{tab:pol_fracs}) and no evidence for circular polarisation. 
Similarly, the PAs for both FRB~121102 \citep{Michilli2018Natur.553..182M} and FRB~190711 do not appear to change as a function of pulse phase. Repeating FRBs have also been largely observed to have wider burst envelopes, with pulse widths ranging from $\sim$ a few ms to 74~ms \citep{chime19R2,CHIME8repeaters1908.03507,CHIME9repeaters2020ApJ...891L...6F} in the 400 to 800 MHz band and $\sim$ a few ms to a few tens of ms in the 1.2 to 8 GHz frequency range \citep{Hessels2019ApJ...876L..23H,Kumar2019ApJ...887L..30K,Marcote2020Natur.577..190M}. Furthermore, \cite{CHIME9repeaters2020ApJ...891L...6F} compared the widths of the repeating and apparently non-repeating FRBs detected by the Canadian Hydrogen Intensity Mapping Experiment (CHIME) and found the repeating FRBs in their sample have larger widths when taken as an ensemble. FRB~190711 is similarly wide with a total burst envelope width of 11.232~ms. We have conducted searches for repetitions with the 64-m Parkes radio telescope as part of an ongoing program to monitor ASKAP-detected FRBs (\citealp{james2020repeat}; Kumar et al., in prep) and have recently identified repetitions from the source (Kumar et al., in prep).

As described in Section \ref{sec:results}, FRB~190711 has three distinct sub-bursts (defined as clearly distinguishable components in both frequency and time). The characteristic frequency (defined as the central frequency of each sub-burst) exhibits a downward drift in frequency with time, as was found for FRB~121102 \citep{Hessels2019ApJ...876L..23H} and for several of the repeat bursts presented in \cite{chime19R2}, \cite{CHIME8repeaters1908.03507}, and \cite{CHIME9repeaters2020ApJ...891L...6F}. As was performed for the 19 repeat bursts of FRB~121102 described in \cite{Hessels2019ApJ...876L..23H}, a drift rate can be determined for the sub-bursts of FRB~190711. \cite{Hessels2019ApJ...876L..23H} found a drift rate range of $\sim$\, 0 to $\rm -865 \, MHz \, ms^{-1}$ using a 2D autocorrelation function analysis. Determination of the drift rate for FRB~190711 is complicated by three factors: there is an intrinsic emission profile, which is drifting downward in frequency with time; this profile has a clear cutoff at higher frequencies that might be intrinsic or extrinsic; and there appears to be a time modulation causing there to be dropouts in the signal. We therefore assume that the bright pixel at roughly 1216~MHz and $\sim$4.0 ms either corresponds to the bright pixel at 1140~MHz and $\sim$8.4 ms or at 1140~MHz and $\sim$9.6 ms. We calculate the drift rate then to be ${\sim}15.4 \pm 1.9 \, \rm MHz \, ms^{-1}$ (where the edges correspond to those edge frequency/time values). This is well within the range of drift rates determined for FRB~121102 \citep{Hessels2019ApJ...876L..23H}.

However, FRB~190711 does show some properties previously unseen or uncommon in repeating FRBs. FRB~190711 has a lower RM than any published repeating FRB (Table~\ref{tab:vela_cal}). FRB~121102 has the highest measured RM of any FRB at $\sim$~$\rm 10^5 \, rad \, m^{-2}$ \citep{Michilli2018Natur.553..182M}, while FRB~180916.J0158+65 has $\rm RM = -114.6 \, rad \, m^{-2}$ \citep{CHIME8repeaters1908.03507}. For FRB~171019, however, \cite{Kumar2019ApJ...887L..30K} found no measurable linear or circular polarisation out to the limit of $\rm |RM| \leq 3 \times 10^4 \, rad \, m^{-2}$ to which they were sensitive.
The existence of the downward drifting frequency-time structure in both FRB~190711 and other repeating FRBs with high RMs illustrates that this feature does not need to originate in a region yielding a high RM.

Along with the apparent temporal modulation, FRB~190711 exhibits frequency modulation (Figure \ref{fig:dynspec}). The significant changes between frequency channels, however, are unresolved by the current channel bandwidth (4~MHz). As the scintillation bandwidth predicted by the NE2001 model \citep{ne2001} is $\gtrsim1.05$~MHz, this frequency modulation could be intrinsic or due to diffractive scintillation, but we are unable to constrain this with the data presented here. We note that the apparent drop in flux seen in Figure \ref{fig:dynspec} (the dark features in Stokes~I just above and below 1200 MHz that persist throughout the pulse) are potentially not physical, as they correspond closely to the regions most heavily contaminated by RFI, and hence the flux calibration is potentially affected in these regions of the spectrum.
\subsection{FRB~190102 and FRB~190611: narrow bursts with time-varying polarisation properties} \label{sec:190102and190611}
FRB~190102 and FRB~190611 both share many phenomenological similarities with FRB~181112 \citep{Cho2020arXiv200212539C}, consisting of multiple narrow components whose polarisation and temporal properties vary. These characteristics are distinct from the properties typically seen in repeating FRBs (i.e., wide bursts with phase-stable polarisation properties) discussed in the preceding subsection. 

The most striking temporal feature is seen in FRB~190611, for which the second sub-pulse exhibits an apparent residual drift in arrival time with frequency (Figure \ref{fig:dynspec}) when de-dispersed using a DM of 332.60 \dmunits, consistent with the optimal value for pulse 1 (332.63 $\pm$ 0.04 \dmunits). A comparable frequency-time drift was seen in pulse 4 of FRB~181112 \citep{Cho2020arXiv200212539C}. As noted in Section \ref{sec:results}, the frequency dependence of this drift is not well constrained, and while well-fitted by a differential dispersion measure, a different origin is plausible. While repeating FRBs have been shown to exhibit a frequency-time drift that is inconsistent with a differential dispersion measure \citep[e.g.][]{Hessels2019ApJ...876L..23H}, this typically results in distinct components drifting across the frequency-time plane, as can be seen in FRB~190711 (see Section \ref{sec:190711} and Figure \ref{fig:dynspec}), rather than a smooth drift in a single component, as seen in the second pulse of FRB~190611. Assuming a $\nu^{-2}$ dependence, the difference in DM between the two pulses is $\rm \Delta DM = 0.26 \pm 0.04 \, pc \, cm^{-3}$, as derived from the analysis described in Section \ref{sec:scatter_method}. This $\Delta \rm DM$ is a factor of $\sim$6 larger than that seen in FRB~181112 (\cite{Cho2020arXiv200212539C}), and as with FRB~181112, the increase in DM for FRB~190611 is observed in the later sub-pulse. Of note, if extrapolated back to infinite frequency, the FRB~190611 sub-pulses would be closer but still temporally separated by $\sim$0.7 ms.

While FRB~181112, FRB~190102, and FRB~190611 all have multiple sub-pulses, the brightest sub-pulse in FRB~190102 is the final one, whereas for FRB~181112 and FRB~190611 the first pulse is the brightest (although the difference in flux density between the two sub-pulses of FRB~190611 is already small and would be further reduced by correcting for the residual drift in the arrival time with frequency.)  Using the Bayesian framework described in Section \ref{sec:scatter_method} and modelling the brightest FRB~190102 sub-pulse and each FRB~190611 sub-pulse with a single Gaussian component convolved with an exponential, we find that the second FRB~190102 sub-pulse and the two FRB~190611 sub-pulses are consistent with being scattered in turbulent plasma (i.e., with a scattering index $\alpha \approx -4$). We note that the low S/N of the initial FRB~190102 sub-pulse precluded a constraining fit with this method. For the main sub-pulse of FRB~190102, we find a scattering time and index of $\rm \tau_{pulse2} = 0.041^{+0.002}_{-0.003} \, ms$ and $\rm \alpha_{pulse2} = -3.84^{+0.71}_{-0.78}$, respectively.
We derive scattering times of $\rm \tau_{pulse1} = 0.18\pm 0.02 \, ms$ and $\rm \tau_{pulse2} = 0.14\pm 0.02 \, ms$ and scattering indices of $\rm \alpha_{pulse1} = -5.86^{+1.73}_{-1.98}$ and $\rm \alpha_{pulse2} = -1.9^{+2.3}_{-2.1}$ for each FRB~190611 sub-pulse. While the precision is lower than in the case of FRB~190102,  due to the lower S/N of the sub-pulses, the derived value for $\alpha$ is consistent between the sub-pulses and consistent with the values derived for the other FRBs presented here. We also determine the intrinsic widths of the FRB~190611 sub-pulses to be $\rm \sigma_{pulse1} = 0.09\pm0.02 \, ms$ and $\rm \sigma_{pulse2} = 0.209\pm0.02 \, ms$. The main FRB~190102 sub-pulse width is $\rm \sigma_{pulse1} = 0.053\pm0.002 \, ms$, where we note that the intrinsic width is consistent with the temporal resolution of the data.

Considering the pulse-averaged polarisation fractions (Table~\ref{tab:pol_fracs}), FRB~190102 and FRB~190611 show many similarities to FRB~181112 \citep{Cho2020arXiv200212539C}. The total polarisation fraction is high in all cases, ranging from $\sim$80\% in FRB~190102 to $>$90\% for FRB~181112 \citep{Cho2020arXiv200212539C} and FRB~190611. However, the polarisation fraction changes between sub-pulses in all cases. FRB~190102 sees only a modest increase in the linear polarisation fraction from the first to second sub-pulse, with a consistent circular polarisation fraction across sub-pulses. The results for FRB~190611, however, are much more striking, with a substantial increase in the circular polarisation fraction while the overall polarisation fraction remains constant. Similar behaviour was seen for pulse 1 and 3 of FRB~181112 \citep{Cho2020arXiv200212539C} and cannot be accounted for via propagation through a cold (i.e., non-relativistic) plasma. This led \cite{Cho2020arXiv200212539C} to conclude that the origins of this change might be in the propagation of the burst through a birefringent medium containing a relativistic plasma, which would lead to generalised Faraday rotation \citep{KennettMelrose1998}. 

FRB~181112, FRB~190102, and FRB~190611 all exhibit a differential RM between pulse components. The magnitude of the RM change is comparable in all cases ($15 \pm 2$, $23 \pm 7$, and $7 \pm 7$ rad m$^{2}$ for FRB~181112, FRB~190102, and FRB~190611, respectively; see \cite{Cho2020arXiv200212539C} and Table~\ref{tab:vela_cal}), but the direction of the change varies: the absolute value of RM increases with time for FRB~181112 \citep{Cho2020arXiv200212539C}, but decreases for FRB~190102 and FRB~190611.  Unlike the time-frequency drift seen in repeating FRBs, which has only been observed to move in one direction (towards lower frequencies with time), this suggests that FRB RMs can vary in either direction.
It is unclear, however, if the difference in RM is the result of propagation along different lines of sight or an intrinsic feature of the emission, or indeed (as noted above) whether the differential RM can be interpreted using an assumption of non-relativistic Faraday rotation. Differential {\em apparent} RMs seen in pulsars have been shown to have no preferred direction of increase \citep{Dai2015MNRAS.449.3223D,Ilie2019MNRAS.483.2778I} and are attributed to processes in the pulsar magnetosphere rather than differential Faraday rotation along the line of sight. 

Of the three FRBs, only FRB~190102 has an RM that is inconsistent with the Galactic contribution estimated along the line of sight to the source: $\rm RM_{MW} = 34 \pm 22 \rm \, rad \, m^{-2}$ \citep{Oppermann2015A&A...575A.118O}. Noting that the predicted $\rm RM_{MW}$ is opposite in sign to our observed RM, the difference of $\sim$150 rad m$^{2}$ could be intrinsic to the source or originate in the intervening material (e.g., the circumburst medium, host ISM, or intervening galaxy halos.) While the sightline to FRB~190102 has not been probed in the same detail as FRB~190608 \citep{Simha+20}, no large galaxies at small impact parameters are present unlike the case of FRB~181112 \citep{Prochaska231}. We therefore conclude it is likely that, as for FRB~190608 \citep{Chittidi2020_HG190608} and FRB~181112, there is likely a substantial contribution to the RM from the host galaxy or local environment of FRB~190102.

The PA swings seen in Figure~\ref{fig:fscrunch} within and between sub-pulses of FRB~190102 and FRB~190611 and Figure 1 in \cite{Cho2020arXiv200212539C} of FRB~181112 further highlight the similarities between these sources and suggest a common emission mechanism. All sources show a more or less bowl-shaped PA curve within each sub-pulse, while FRB~181112 and FRB~190102 also show a significant difference in the mean PA between pulses (with $\rm \Delta\Psi_{mean} \sim 20 \degree$).
As discussed in \cite{Cho2020arXiv200212539C}, the evolution in PA across the FRB can be used to distinguish between geometric configurations of the emission region. If these variations are due to an intrinsic magnetic field reconfiguration, this would require significant topological changes to occur on sub-ms timescales.
If, however, pulsar-like emission is assumed, in which the emission sweeps across the sightline, a static or slowly varying magnetic field can account for the variable PA. Following \cite{Cho2020arXiv200212539C}, we calculate the minimum spin period for a putative rotating source assuming a rotating vector model for the polarisation position angle as a function of time. The maximum measured change of 55 degrees per millisecond for FRB~190102 and 70 degrees per millisecond for FRB~190611 yields a lower limit on the putative spin periods of
\begin{eqnarray} \label{eq:period}
P_{\mathrm{FRB} 190102} > 5.1\,{\rm ms} \, \left\vert \frac{\sin \alpha}{\sin \beta} \right\vert  \\
P_{\mathrm{FRB} 190611} > 6.4\,{\rm ms} \, \left\vert \frac{\sin \alpha}{\sin \beta} \right\vert  ,
\end{eqnarray}
where $\alpha$ and $\beta$ are defined in \cite{Cho2020arXiv200212539C} as the angles between the spin axis and magnetic dipole axis and the magnetic dipole axis and the sightline, respectively. The differing PA curves in pulse 1 and pulse 3 of FRB~181112 led \citet{Cho2020arXiv200212539C} to argue against all four sub-pulses being emitted within a single rotation, if rotation were assumed for the source. However, given the similarity in the PA curves for the two FRB~190611 sub-pulses, it is plausible that these might be successive views of the same emission region one rotation later. That is, the intrinsic spin period could be $\sim 1 \, \rm ms$ if interpreted in this way. The significant change in the polarisation fractions between the sub-pulses argues against this interpretation, however, as does the fact that FRB~181112 and FRB~190102 have multiple components with similar temporal separations that cannot be interpreted this way.

The FRB~190611 dynamic spectra (Figure \ref{fig:dynspec}) clearly reveal frequency banding on two scales. The bright, narrow frequency structure within each sub-pulse appears strongly correlated between the two sub-pulses, while the overall emission envelope appears to shift between sub-pulses, with the second sub-pulse peaking at a higher frequency than the first. In order to determine the level of correlation between both sub-pulses and between the fine-scale structure within each sub-pulse, a cross correlation function (CCF) and an autocorrelation function (ACF) were respectively used to obtain lag spectra between bins 15-16 (pulse 1) and 24-26 (pulse 2) and for each individual sub-pulse. At the 4~MHz resolution of our data, we do not resolve the small-scale modulation in the ACF data for either sub-pulse. This is consistent with predictions of the diffractive scintillation bandwidth of $\gtrsim1.00$~MHz predicted by the NE2001 model \citep{ne2001}, and thus the small-scale modulation is plausibly explained by diffractive scintillation. In the CCF function, in addition to a narrow peak at zero offset, a broad peak is seen at an offset of $-48$~MHz (i.e., shifting the second sub-pulse 48~MHz lower in frequency), providing evidence that the overall emission envelope as a function of frequency differs between the two sub-pulses. This cannot be ascribed to diffractive scintillation, and thus, we conclude that this is related to the intrinsic emission mechanism.

While the FRB~190102 dynamic spectra (Figure \ref{fig:dynspec}) do not exhibit significant spectral features, they do show clear inter-channel variations in intensity that are inconsistent with thermal noise and do not evolve strongly with frequency. The NE2001 \citep{ne2001} prediction for the scintillation bandwidth is $\gtrsim1.02$~MHz, and hence the effects of Galactic diffractive scintillation may be obscured by our 4-MHz channel resolution. To determine if the modulation is likely due to diffractive scintillation, we calculate the cumulative distribution function (CDF) of the intensities for both sub-pulses separately and fit each with an exponential distribution, which is the expected distribution in the case of diffractive scintillation. We find that neither sub-pulse is well fit by an exponential, favouring an intrinsic mechanism as the source of the frequency modulation. 

The circular polarisation component in the second sub-pulse of FRB~190102 has a curious appearance, exhibiting frequency-dependent structure (e.g., the sign change in Stokes~V seen in Figure \ref{fig:dynspec}). However, we consider this most likely a residual calibration error, noting that the magnitude of the Stokes~V component is only a few percent of the (very bright) linearly polarised emission. As described in Section \ref{sec:pol_cal}, the polarisation calibration technique used is a linear approximation rather than a true bandpass calibration, meaning deviations from this linear approximation leakage will result in leakage. Given these limitations in our polarisation calibration, we treat this apparent low-level structure in Stokes~V with caution.

\subsection{FRB~180924 and FRB~190608: substructure obscured by scattering} \label{sec:180924and190608}

The detection and localisation of FRB~180924, along with its host galaxy properties and a limited analysis of its time domain properties were reported in \citet{Bannister565}. Based on the higher time resolution analysis performed here, we update the DM of FRB~180924 \citep[previously reported to be $\rm 361.42 \pm 0.06 \, pc \, cm^{-3}$;][]{Bannister565}, as shown in Table~\ref{tab:frb_props}.

When fit with a single Gaussian component, FRB~180924 yields a narrow component width $\sim$0.1\,ms, comparable to the widths seen for FRB~190102 and FRB~190611, while FRB~190608 is considerably wider at $\sim$1.1\,ms (Table~\ref{tab:frb_props}). These two FRBs are the most heavily scattered of our sample, with a scattering time of 0.68 and 3.3~ms, respectively. However, the frequency-averaged pulse profiles of both FRB~180924 and FRB~190608 hint at the existence of multiple components blended into the scattering tail of the first, brightest component (Figure \ref{fig:fscrunch}). As an initial step in evaluating the existence of multiple pulses in FRB~180924, a set of four sub-banded, frequency-averaged time series were made from the dynamic spectra and inspected, showing no significant difference in the arrival times of any components.

\begin{figure*}
    \centering
    \includegraphics[width=0.45\textwidth]{{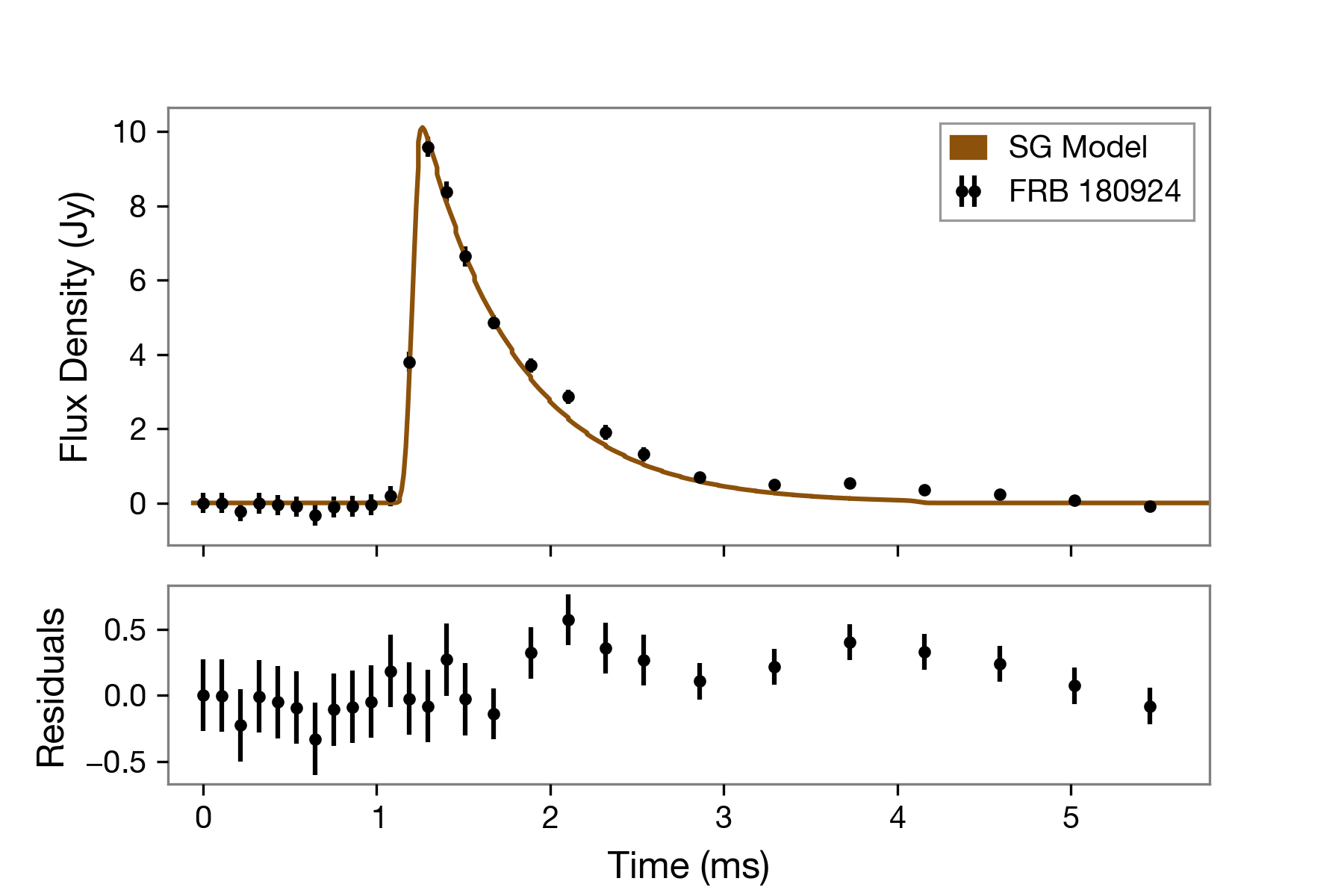}}
    \includegraphics[width=0.45\textwidth]{{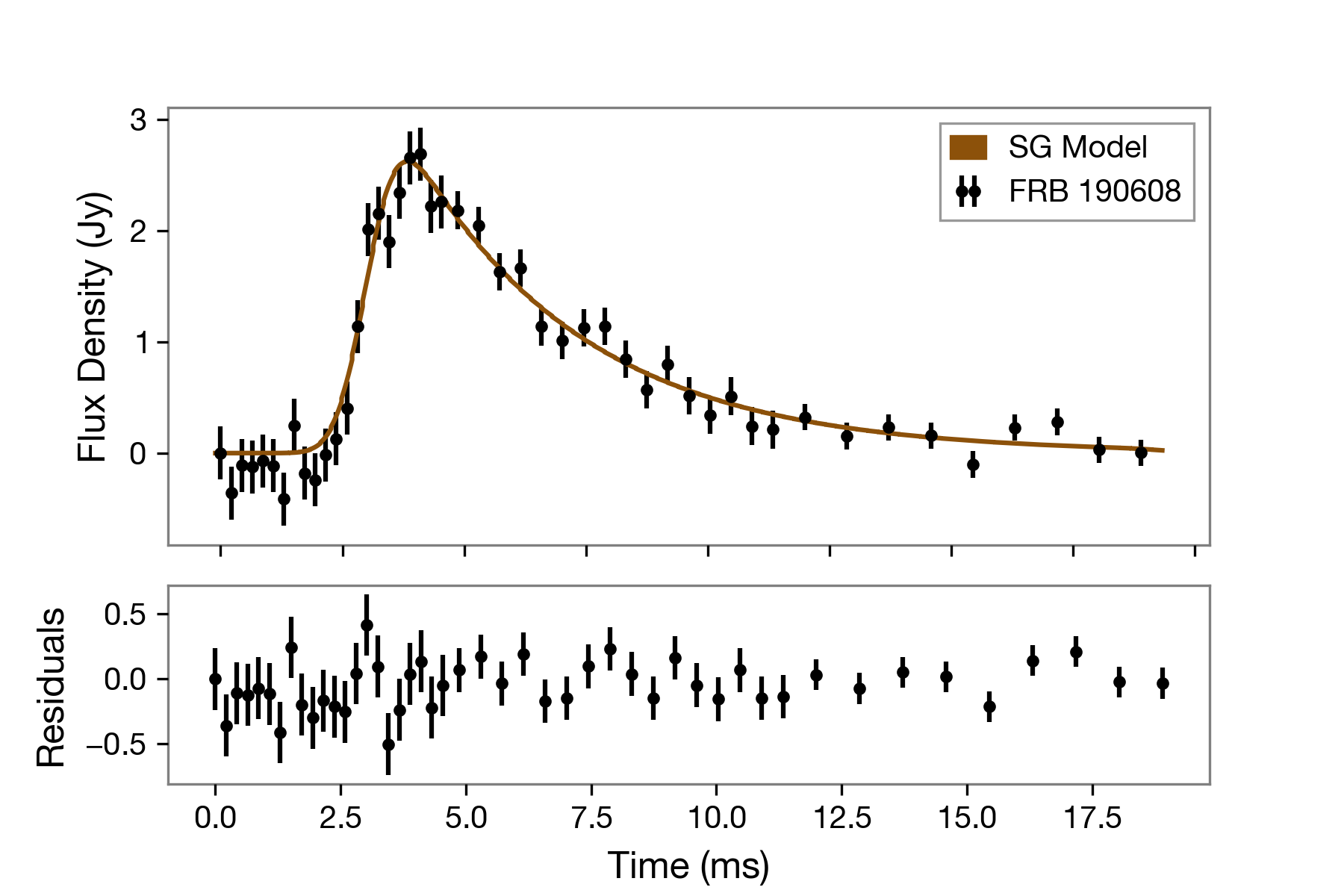}}\\
    \includegraphics[width=0.45\textwidth]{{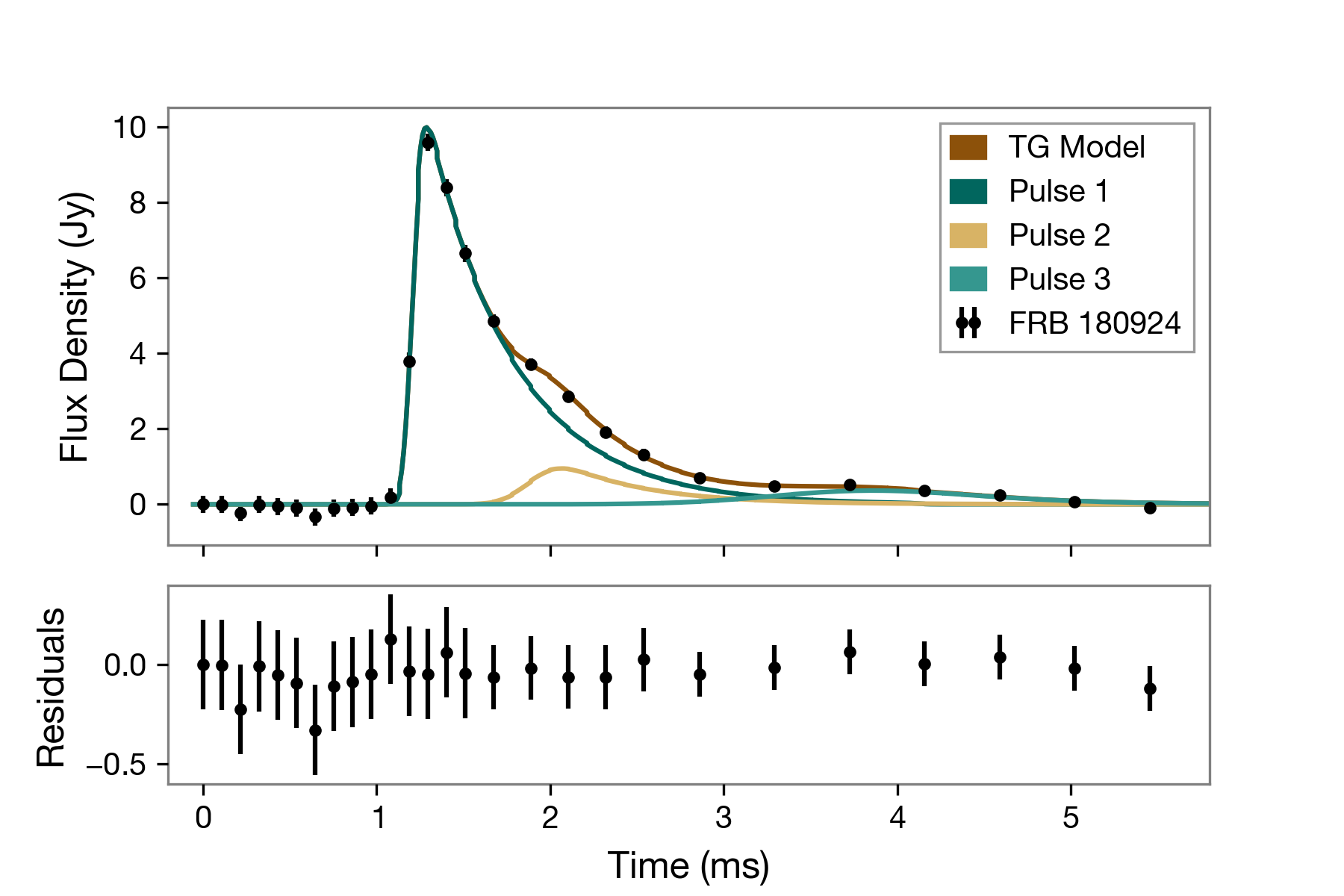}}
    \includegraphics[width=0.45\textwidth]{{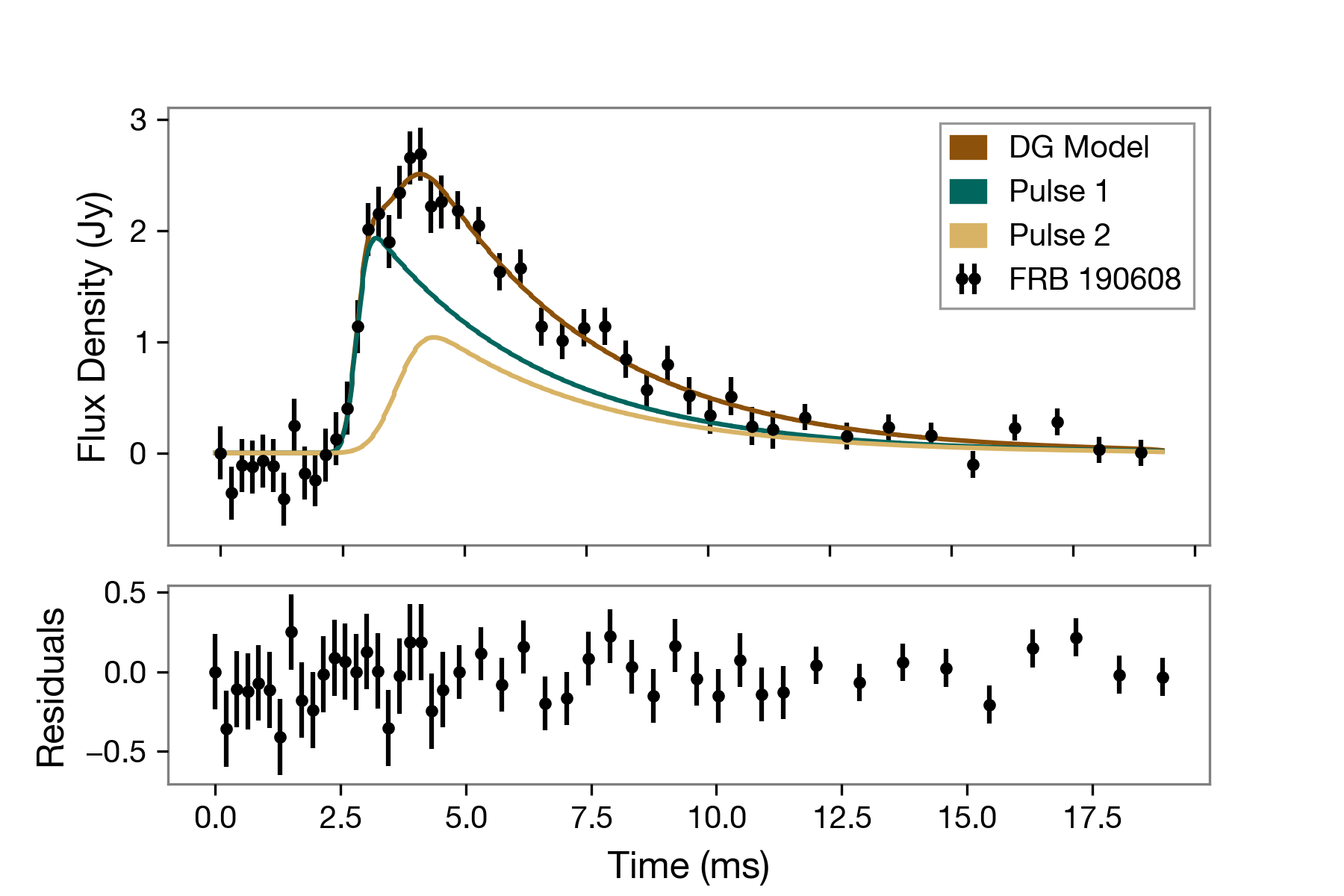}}\\
    \caption{Stokes~I time series for
    FRB~180924 with a single pulse model fit (upper-left), FRB~180924 with a three component model fit (lower-left), FRB~190608 with a single pulse model fit (upper-right), and FRB~190608 with a two component model fit (lower-right).
    The best fit models for each FRB are plotted over the data with residuals displayed in the bottom panel. For the multiple component models, we also display the pulse components separately to highlight the location of the pulses. For FRB~190608, the single wide pulse cannot represent the rapid rise time adequately, as can be seen in the residuals. For display purposes only, we have averaged the lower S/N data in two sections of each time series. For FRB~180924, the ranges 1.6 -- 2.7~ms and 2.7 -- 5.7~ms were averaged by a factor of 2 and 4, respectively, and for FRB~190608, the ranges 4.8 -- 11.7~ms and 11.7 -- 19.2~ms were averaged by a factor of 2 and 4, respectively.}
    \label{fig:multicomponents}
\end{figure*}

In order to further investigate the scattering-obscured structure and characterise the properties of any additional components in FRB~180924 and FRB~190608, we considered a multi-component model and compared the Bayesian evidence over the single-component model, as described in Section \ref{sec:scatter_method}, for both FRBs. For FRB~180924, the results show strong evidence ($\Delta \mathrm{LogE} \sim134$) for two fainter and wider ($\sigma_2 \rm < 0.4 \,ms$ and $\sigma_3 \rm = 1.0^{+0.5}_{-0.4} \,ms$, respectively) components offset by $\rm 0.68 \,ms$ and $\rm 2.35 \,ms$, respectively, from the first ($\sigma_1 \rm \sim 0.06 \pm 0.02 \,ms$). We note that, while the width of the second component is an upper limit (i.e., unresolved at the current data resolution), the three-component model is favoured over a two-component model since the former provides an improved fit to both the ``shoulder'' (at $\sim$2~ms) and the low-level broad emission beyond 3~ms. We display the three Gaussian + scattering components and the combined model fit in the lower left panel of Figure \ref{fig:multicomponents}. Such broad, late-time emission as modelled by component three has not been noted in previous FRB detections, but it would have been difficult or impossible to discern at lower signal-to-noise ratios. The S/N boost in our data relative to the initial detection (facilitated by the retention of the ASKAP voltage data), however, enables this to be observed. Further examples of high S/N bursts, which will be common with ASKAP, could confirm whether this feature is ubiquitous.

For FRB~190608, the model comparison favours two moderately broad ($\sigma_1 \rm = 0.3 \pm 0.1 \,ms$ and $\sigma_2 \rm = 0.6 \pm 0.4 \,ms$, respectively) components over the single-component model ($\Delta \mathrm{LogE} \sim 56$), where the second component is offset by $\rm 0.82 \,ms$ from the first. Figure~\ref{fig:multicomponents} shows the best fitting single Gaussian + scattering model, along with the residuals, in the top right panel and the two-component model and residuals in the lower right panel. The clearest discrepancies in the single-component model are around the rising edge of the pulse, where the wide single component is unable to reproduce the relatively sharp rise. The addition of a second component, however, better captures the rapid rise time. We note that the two averaged points in the range 16 -- 17.5~ms appear to be above the scattering tail in both models, which may indicate a third, broad component, as seen in FRB~180924. However, the relatively low overall S/N of this FRB makes fitting weaker components in individual sub-bands and hence constraining the properties of additional sub-pulses difficult.  Alternate approaches that apply tighter priors on, e.g., differential dispersion between sub-pulses may be able to better characterise weaker components in a future analysis.

The pulse-averaged polarisation properties of FRB~180924 and FRB~190608 are nearly identical -- each has $\sim$90\% linear polarisation\footnote{We note that \citet{Bannister565} reported a linear polarisation fraction of $80 \pm 10$\% for FRB~180924, which we update here using the higher resolution data and improved calibration.} and $\sim$10\% circular polarisation (Table~\ref{tab:pol_fracs}). The polarisation position angle behaviour, however, differs substantially. The PA of FRB~180924 is flat in time, resembling that of FRB~190711, while FRB~190608 shows a marked, near-linear drift with time.  While flattening of a pulse PA can be attributed to scattering \citep[e.g.,][]{Caleb2018MNRAS.478.2046C,Li&Han2003A&A...410..253L}, this does not typically result in a linear change in the PA. In the case of multiple scattered components, however, the overall PA behaviour would depend on the separation, amplitude, and PA of the individual components as well as the scattering timescale. Components with comparable PA would lead to a flat PA throughout the scattered pulse (FRB~180924 is not overly dissimilar to how FRB~190611 would appear after experiencing comparable scattering), but components with distinct PA values (like FRB~190102, albeit with considerably different flux density ratios and widths) could be blurred together and generate a monotonic PA trend.

Both FRB~180924 and FRB~190608 have frequency structure (Figure \ref{fig:dynspec}) that may be consistent with diffractive scintillation. This is unrelated to the large scattering observed for these two FRBs, which would manifest as scintillation with bandwidths $<1$~kHz given the ms-level scattering times, and would instead require the presence of a second (Galactic) scattering screen. The NE2001 \citep{ne2001} prediction for the scintillation bandwidth is comparable for each sightline, at $\gtrsim2.2$ and $\gtrsim2.4$~MHz, respectively, meaning the decorrelation bandwidth may fall below the resolution of our data. For each FRB, we calculate a frequency ACF using a slice of the data that roughly spans the half-power points of the pulse. For FRB~190608, we find no significant peaks in the lag spectrum beyond the zeroth lag and accordingly are unable to confirm if the origin of this frequency banding is diffractive scintillation with the current data resolution. 

For FRB~180924, we fitted a Lorentzian function to the ACF lag spectrum, following \citet{Cho2020arXiv200212539C}, and confirm that the decorrelation bandwidth is 8.5~MHz, as reported by \citet{Bannister565}. Moreover, following the method used for FRB~190102 (Section \ref{sec:190102and190611}), we calculate the CDF of the intensities and fit this with an exponential distribution, finding that this describes the data well, further suggesting diffractive scintillation as the origin of the frequency structure. While we cannot rule out intrinsic spectral structure in FRB~180924, the large-scale structure observed at the current resolution is consistent with diffractive scintillation.

The FRB~180924 RM ($\rm 22 \pm 2 \, rad \, m^{-2}$) is similar to those of FRB~190611 and FRB~190711 and broadly consistent with the estimated Milky Way contribution (Table~\ref{tab:vela_cal}). We note that the high resolution data has enabled an improved derivation of the RM over the previously reported $\rm RM = 14 \pm 1 \rm \, rad \, m^{-2}$ \citep{Bannister565}. FRB~190608, on the other hand, has the highest RM of the sample presented in this paper, with $\rm 353 \pm 2 \, rad \, m^{-2}$, a value that considerably exceeds the expected Milky Way contribution and suggests a substantial contribution from the host environment.  The properties of both the host galaxy of FRB~190608 and its foreground were respectively studied extensively in \citet{Chittidi2020_HG190608} and \citet{Simha+20}. Using the foreground halo contribution estimation of $< 1 \rm \, rad \, m^{-2}$ from \citet{Simha+20}, \citet{Chittidi2020_HG190608} concluded the bulk of the excess RM originated within the host, likely containing contributions from both the host ISM and the local environment. 

\citet{Chittidi2020_HG190608} also investigated the possible origins of the scatter broadening of FRB~190608, finding it could not be fully explained via scattering in the ISMs of either the Milky Way or host. \citet{Simha+20} estimated a negligible contribution from intervening turbulent material along the line of sight, and \citet{Chittidi2020_HG190608} argued that two scenarios were therefore plausible for the origin of the large scattering timescale: (1) a highly dense, turbulent material very close to the source or (2) a highly turbulent, dense H\,{\sc ii} region along the sightline within the host. Considering the measured decorrelation bandwidth of FRB~180924, which yields a scattering time $\sim 0.01 \, \mu \rm s$ from the Milky Way (note that \cite{ne2001} predict a value $\gtrsim 0.05 \, \mu s$), the host galaxy is the more likely origin of the ms-scale scattering seen in FRB~180924. Studies similar to those conducted for FRB~190608 by \citet{Chittidi2020_HG190608} and \citet{Simha+20} are necessary, however, to constrain the location of the scattering for FRB~180924 and are presently underway (Simha et al., in prep).

Overall, we conclude that the underlying structure of FRB~180924 and FRB~190608 share many similarities to FRB~181112, FRB~190102, and FRB~190611, despite initially appearing to be a wider, single-component burst -- largely because of the stronger scattering seen in these bursts. However, the third component of FRB~180924 would be the widest of any of the sub-pulses of any of the bursts clearly within the category typified by FRB~181112 by a factor of $\sim 5$ (although only a factor of $\sim 2$ wider than the widest FRB~190608 sub-pulse). While FRB~190608 exhibits the highest RM in our sample and the largest degree of scattering, both of which can be explained by a dense and magnetised circumburst medium favoured for some repeating FRB models, the non-zero circular polarisation and time-varying polarisation position angle do not fit the (admittedly poorly constrained) repeater archetype and could adequately be explained by the favoured multi-component model, in which the individual components are heavily blended by scattering. A detected repeat from either source (or strong limits against detection) would enable further constraints on the characteristics of repeating (or apparently non-repeating) FRBs.
\section{CONCLUSIONS} \label{sec:conclusions}

We have presented the high time and spectral resolution, full polarisation analysis of five localised ASKAP FRBs with exceptionally high signal-to-noise ratios and investigated their properties. We find that scattering is detected in all cases for which a fit could be obtained -- noting that the complex temporal and spectral structure of FRB~190711 precludes fitting a scattering model -- with a mean scattering index of $-3.7 \pm 0.4$, consistent with scattering caused by turbulent plasma \cite[][]{bhatscattering}. We find in each case that the scattering time is inconsistent with predictions based on models of the Galactic electron density distribution and conclude that those FRBs with detectable scattering are scattered outside the Milky Way. The required scattering screens may be found local to the source, within the host galaxy, within the IGM, or within any intervening galaxies along the line of sight. In the case of FRB~190608, the host galaxy and foreground analyses conducted by \citet{Chittidi2020_HG190608} and \citet{Simha+20}, respectively, indicate that the scattering is likely originating from within the host galaxy (either from the ISM or the source-local material). Similar future studies would constrain the origins of the scattering for FRB~180924, FRB~190102, and FRB~190611. If the scattering is generated near the FRB source in most cases, we cannot immediately relate the strength of the scattering to any property of the host galaxy or local environment. The fitted scattering widths to our sample of FRBs spans a wider range (two orders of magnitude) than the host galaxy masses or star formation rates \citep{Bhandari2020_hostgalaxies}. It is also noteworthy that the two most strongly scattered FRBs in our sample, FRB~180924 and FRB~190608, originate in the outer environs of their host galaxies \citep{Macquart2020_DMz,Chittidi2020_HG190608}, implying the host ISM is not the first order origin of the scattering but rather the circumburst medium. In this scenario, any source-local scattering medium must also satisfy the requirement for a wide range of local RM contributions.

There is strong evidence that all FRBs within our sample have multiple components. FRB~190102 and FRB~190611 have multiple, distinct narrow components similar to FRB~181112 \citep{Cho2020arXiv200212539C}, and FRB~190711 has clear sub-burst structure. The pulse profiles of FRB~180924 and FRB~190608 show evidence for temporal substructure obscured by scattering of the leading component. A three-component scattered Gaussian model, which includes broad extended emission at late times, is clearly preferred over a single scattered Gaussian model for FRB~180924. Likewise, a two-component scattered Gaussian model is favoured over a single-component model for FRB~190608. The scattering time of FRB~190608 is a factor of $\sim 10$ greater than that of FRB~180924, however, which possibly acts to mask a third, faint component. As the PA values associated with the broad, late-time emission seen in FRB~180924 (and posited for FRB~190608) are both consistent with the preceding PAs (or consistent with the PA trend, in the case of FRB~190608) and lie beyond the scattering tail of the brightest component in each FRB, this argues for at least one additional, faint component. This coupled with the evolving PA within the main scattering region of the pulse profile, which is most naturally explained via multiple components, offer a strong case for their existence.

Although there is some evidence for emerging sub-classes within our sample of five FRBs, we find no clear distinction between bursts that appear consistent with the canonical ``repeating'' and apparently non-repeating FRBs. Rather, our sample appears to form a continuous spectrum of features bridging the potential divide between the two often proposed populations. As discussed in Section \ref{sec:190711}, FRB~190711 -- the sole known repeater in our sample (Kumar et al., in prep) -- exhibits many of the characteristic features associated with repeating FRBs, namely the downward drifting time-frequency structure (Figure \ref{fig:dynspec}), a wide burst envelope (Figure \ref{fig:fscrunch}), a linear polarisation fraction consistent with 100\% with negligible circular polarisation (Table~\ref{tab:pol_fracs}), and a flat PA (Figure \ref{fig:fscrunch}). FRB~190102 and FRB~190611, conversely, appear to be consistent with a distinct category to which FRB~181112 also belongs \citep{Cho2020arXiv200212539C}: they contain multiple narrow sub-pulses, have significant circular polarisation fractions, exhibit PA swings and changing polarisation fractions, and lack the typical downward drift of a repeater. The categorisation of FRB~180924 and FRB~190608 is made more challenging by their larger scattering timescales. While FRB~180924 initially appears to have some repeater-like characteristics -- high linear polarisation with a flat PA -- closer inspection reveals evidence for multiple, narrow components with moderate circular polarisation more akin to FRB~181112-like bursts, where scatter-broadening has yielded a long flat PA. Similarly, FRB~190608 shares some features often associated with repeating FRBs, including a high linear polarisation fraction. In addition, its relatively high RM could arise from a dense, magnetised medium local to the source, an environment favoured for many repeating FRB models. However, it also has a moderate circular polarisation fraction and a variable PA. As with FRB~180924, a plausible origin of the PA variations is the existence of multiple scattered components.

Along with the temporal and spectral features, the polarisation properties of the FRBs in our sample yield clues to the environments of their sources. FRB~190711 has the lowest measured RM of any repeater, indicating that repeating FRBs need not originate in regions associated with strong, ordered magnetic fields. This range in possible RM magnitudes for repeating FRBs suggests that RMs cannot be used deterministically to associate FRBs with any hypothesised repeating versus non-repeating class. Likewise, the range in RMs within our sample, including within the FRB~181112-like FRBs, illustrates that their common features do not necessitate regions with similar magnetic field strengths or topology. While FRB~180924, FRB~190611, and FRB~190711 have RMs consistent with the predicted Galactic contribution \citep{Oppermann2015A&A...575A.118O}, FRB~190102 and FRB~190608 have RMs significantly in excess of the Galactic contributions. \citet{Chittidi2020_HG190608} concluded the large excess FRB~190608 RM likely originated within the host galaxy. While a more complete study is required to better constrain the host or intrinsic contribution to the FRB~190102 RM, a substantial intrinsic or local/host contribution cannot be excluded. Additionally, the apparent exchange of linear to circular polarisation has been observed in multiple FRBs (e.g., FRB~181112, FRB~190102, and FRB~190611); thus, it is imperative that future models are capable of explaining this behaviour. We also note that the majority of our FRBs are nearly 100\% polarised. Current FRB progenitor models \citep[e.g., ][]{Margalit2020} predict high linear polarisation fractions. Likewise, magnetars, which are known to exhibit high linear polarisation fractions \citep[e.g., ][]{Levin2012,Shannon2013,Lower2020}, are often invoked in the source models of FRBs \citep[e.g., ][]{Margalit2019,Metzger2019MNRAS.485.4091M}. We note that natural sources of nearly 100\% polarised emission are rare, and our sample provides stronger constraints for the prevalence of high total polarisation fractions as well as further evidence of both a high fractional and variable circular polarisation component in at least a subset of FRBs.

\section*{Acknowledgements}

C.K.D. acknowledges the support of the CSIRO Postgraduate Scholarship - Astronomy and Space (47417).
A.T.D. is the recipient of an ARC Future Fellowship (FT150100415).
R.M.S. is the recipient of an ARC Future Fellowship (FT190100155).
R.M.S., K.B., and J.-P.M. acknowledge Australian Research Council (ARC) grant DP180100857.
H.Q. acknowledges the support of 
the Hunstead Merit Award for Astrophysics from the University of Sydney.
J.X.P.,
as member of the Fast and Fortunate for FRB
Follow-up team, acknowledges support from 
NSF grants AST-1911140 and AST-1910471.
The Australian SKA Pathfinder is part of the Australia Telescope National Facility which is managed by CSIRO. Operation of ASKAP is funded by the Australian Government with support from the National Collaborative Research Infrastructure Strategy. ASKAP uses the resources of the Pawsey Supercomputing Centre. Establishment of ASKAP, the Murchison Radio-astronomy Observatory and the Pawsey Supercomputing Centre are initiatives of the Australian Government, with support from the Government of Western Australia and the Science and Industry Endowment Fund. We acknowledge the Wajarri Yamatji people as the traditional
owners of the Observatory site.

\section*{Data Availability}
The data underlying this article will be shared on reasonable request to the corresponding author.





\bibliographystyle{mnras}
\bibliography{htr+pol}






\bsp	
\label{lastpage}
\end{document}